\documentclass[preprint,nofootinbib,superscriptaddress,12pt]{revtex4}
\usepackage{amsmath,amssymb,graphicx,latexsym,slashed}

\usepackage[hypertex]{hyperref}

\newcommand{\be}{\begin{equation}}
\newcommand{\ee}{\end{equation}}

\def\pcmatrix#1{\begin{pmatrix} #1 \end{pmatrix}}

\newcommand{\vev}[1]{\left\langle #1 \right\rangle}

\newcommand{\eps}{ \epsilon }

\newcommand{\tgb}{ t_\beta }

\newcommand{\bea}{\begin{eqnarray}}
\newcommand{\eea}{\end{eqnarray}}

\renewcommand{\Re}{\mathcal{R}e}
\renewcommand{\Im}{\mathcal{I}m}

\newcommand{\beq}{\begin{equation}}
\newcommand{\eeq}{\end{equation}}
\newcommand{\beqa}{\begin{eqnarray}}
\newcommand{\eeqa}{\end{eqnarray}}

\newcommand{\lsim}{\mathrel{\rlap{\lower4pt\hbox{\hskip1pt$\sim$}}
    \raise1pt\hbox{$<$}}}         
\newcommand{\gsim}{\mathrel{\rlap{\lower4pt\hbox{\hskip1pt$\sim$}}
    \raise1pt\hbox{$>$}}}         

\begin{document}

\vspace*{.0cm}

\title{Non-Standard Neutrino Interactions at One Loop}

\def\cornell{Institute for High Energy Phenomenology\\
Newman Laboratory of Elementary Particle Physics\\ Cornell University,
Ithaca, NY 14853, USA}
\def\PPaddr{Physik-Department, Technische Universit\"at M\"unchen, D-85748 Garching, Germany\vspace{2cm}}

\author{Brando Bellazzini}\email{b.bellazzini@cornell.edu}\affiliation{\cornell}
\author{Yuval Grossman}\email{yg73@cornell.edu}\affiliation{\cornell}
\author{Itay Nachshon}\email{in36@cornell.edu} \affiliation{\cornell}
\author{Paride Paradisi}\email{paride.paradisi@ph.tum.de}\affiliation{\PPaddr}



\begin{abstract}
Neutrino oscillation experiments are known to be sensitive to
Non-Standard Interactions (NSIs). We extend the NSI formalism to
include one-loop effects. We discuss universal effects induced
by corrections to the tree level $W$ exchange, as well as
non-universal effects that can arise from scalar charged current
interactions.
We show how the parameters that can be extracted from the experiments
are obtained from various loop amplitudes, which include vertex
corrections, wave function renormalizations, mass corrections
as well as box diagrams. As an illustrative example, we discuss 
NSIs at one loop in the Minimal Supersymmetric Standard Model (MSSM) 
with generic lepton flavor violating sources in the soft sector. 
We argue that the size of one-loop NSIs can be large enough to be 
probed in future neutrino oscillation experiments.
\end{abstract}

\maketitle

\section{Introduction}

There are many experimental results confirming that neutrinos have
masses and oscillate between different flavors~\cite{review}. It is
plausible that neutrinos acquire small masses from some high scale
physics via the see-saw mechanism~\cite{seesaw}. If such high energy
dynamics are the only source of lepton flavor violation (LFV),
neutrino oscillations could remain their only observable
effects.\footnote{A remarkable exception arises in the context of
supersymmetric theories where such high scale dynamics could leave
indelible footprints on the soft terms of the light sparticles via
interactions not suppressed by inverse powers of the high
scale~\cite{Hall:1985dx,Borzumati:1986qx}.} It is possible, however,
that extra sources of LFV are present at the weak scale. Such
LFV sources could induce lepton number violating decays of charged
leptons such as $\mu\to e\gamma$, $\mu\to eee$ and $\tau\to\mu\gamma$.
Weak scale LFV interactions can also affect neutrino oscillation
experiments. For instance, flavor violating Non-Standard Interactions
(NSIs) can modify neutrino propagation in
matter~\cite{Wolfenstein:1977ue,Mikheev-Smirnov,Roulet:1991sm,Guzzo:1991hi,NewPhysMatter,Fornengo:2001pm}.

NSIs could also affect the production and detection processes
generating ``wrong flavor''
neutrinos~\cite{Grossman:1995wx,GonzalezGarcia:2001mp,Ota:2001pw}.
Consider for instance an appearance experiment where a neutrino is
produced in association with a muon, and a tau is detected at the
detector. The standard interpretation of such a result is that it is
due to $\nu_\mu\to\nu_\tau$ oscillations.  This interpretation is
correct as long as the produced neutrino is orthogonal to the detected
one. However, if there are NSIs that violate the flavor symmetry at
the source, the produced neutrino, which we denote by $|\nu_\mu^s\rangle$
is not a simple flavor eigenstate. Analogously, in the presence of
detector NSIs, the final state, which we denote as $|\nu_\tau^d\rangle$,
is not a simple flavor eigenstate. If these states are not orthogonal, 
that is,
\beq
\vev{\nu_\mu^s|\nu_\tau^d}\equiv\varepsilon_{\mu \tau} \ne 0~,
\eeq
tau appearance can occur without oscillation.

In order to describe the effect of NSIs on neutrino oscillation experiments,
let us discuss first a simple case with only two generations and a mixing angle
$\theta=\pi/4$. Consider also the case when the detector is at a distance $L$
much smaller than one oscillation length,
\beq
 x\equiv{\Delta m^2 L \over 4 E} \ll 1~.
\eeq
In this case the oscillation amplitude simply reads ${\cal A}_{\rm
osc}(\nu_\mu\to\nu_\tau)\approx i x$. When NSIs are present, they
induce an extra contribution, ${\cal A}_{\rm
NSI}(\nu_\mu\to\nu_\tau)=\varepsilon_{\mu\tau}$. Thus, to leading
order in $x$, the total appearance probability is given by the squared
of the sum of amplitudes
\beq
\label{gen-eq}
{\cal P}(\nu_\mu \to \nu_\tau) \approx \left|{\cal A}_{\rm osc}+{\cal
A}_{\rm NSI}\right|^2
\approx
x^2 + |\varepsilon_{\mu \tau}|^2 + 2 x \; \Im\left(\varepsilon_{\mu
\tau}\right)~.
\eeq
The above simplified result has all the physics in it: the first term
is the pure oscillation term, the second one is the $x$ independent
term that arises due to the NSIs, and the third term is an
interference term. Note that, when $x\gg\varepsilon_{\mu\tau}$, the
probability to detect a new physics effect is \textit{enhanced} by the
interference term, which is linear in $\varepsilon_{\mu\tau}$,
compared to the typical quadratic dependence of lepton flavor
violating decays of charged leptons~\cite{GonzalezGarcia:2001mp}. This
fact triggered renewed interest at present neutrino
facilities~\cite{Ota:2002na,Friedland:2005vy,Kitazawa:2006iq,Friedland:2006pi,Blennow:2007pu,Blennow:2008ym}. The
typical bounds on production and detection NSIs are about
$\varepsilon_{\mu\tau} \sim
10^{-2}$~\cite{Biggio:2009kv,Biggio:2009nt}. Higher sensitivities, of
$\varepsilon_{\mu\tau} \lsim 10^{-3}$ are within the reach of future
neutrino
experiments~\cite{Ohlsson:2008gx,Kopp:2007mi,Kopp:2007ne,Ribeiro:2007ud,Holeczek:2007kk,Goswami:2008mi,Antusch:2009pm,Altarelli:2008yr,Bandyopadhyay:2007kx}.

In this paper we study NSIs at the loop level, extending the formalism
of~\cite{Grossman:1995wx,GonzalezGarcia:2001mp}. We present a general
framework that allows one to extract in a consistent way the physical
parameters $\varepsilon_{\alpha\beta}$ (with $\alpha,\beta=1,3$) which
arise at the loop level either from corrections to the tree level $W$
exchange diagram or from more general corrections, in particular from
scalar charged currents.  We show how the physical parameter,
$\varepsilon_{\alpha\beta}$, can be obtained from the various loop
amplitudes which include vertex corrections, wave function
renormalizations, mass corrections as well as box diagrams.

In the case of universal corrections to the $W$ exchange amplitudes,
NSIs emerge at one loop because after the Electroweak Symmetry
Breaking (EWSB) the kinetic terms and the $W$ couplings are generally
not universal in the same basis. Rotating to the mass basis for the
charged leptons, the misalignment between vertex and wave functions
induce NSIs.  We show that the associated $\varepsilon_{\alpha\beta}$
are finite because the $SU(2)_L$ gauge symmetry protects them from
possibly divergent contributions.

This paper is organized as follows. In Sec.~\ref{Formalism}, we recall
the standard formalism for NSIs. In Sec.~\ref{Oneloop}, we extend it
to account for loop induced NSIs. In Sec.~\ref{nsi_susy}, we discuss
NSIs at one loop in the R-parity conserving Minimal Supersymmetric
Standard Model (MSSM) with generic LFV sources in the soft sector,
leaving the details of the calculation to the appendix. Finally, we
present our conclusions in Sec.~\ref{conclusions}.

\section{Notations and formalism}
\label{Formalism}

We start by setting our notations to follow
Refs.~\cite{Grossman:1995wx,GonzalezGarcia:2001mp}.  We first consider
only tree level processes and later discuss one loop effects.

Neutrino mass eigenstates are denoted by $|\nu_i\rangle$, $i=1,2,3$,
while $|\nu_\alpha\rangle$ are the tree level weak interaction
partners of the charged lepton mass eigenstates $\alpha^-$,
$\alpha=e,\mu,\tau$.  These two bases are related by
\beq
\label{nuellW}
|\nu_\alpha\rangle=\sum_i U_{\alpha i}|\nu_i\rangle~.
\eeq
where $U$ is the leptonic mixing matrix, the so-called PMNS
matrix~\cite{Pontecorvo:1957cp}.

We consider experiments where neutrinos are produced at the source
in conjunction with incoming negative or outgoing positive charged
leptons. Then, the neutrinos travel to the detector where they are
detected by producing negative charged leptons. We consider new
physics in the production and detection processes assuming that these
NSIs have the same Dirac structure as the SM interactions and thus
the amplitudes add coherently. We parameterize the new interactions
at the source and at the detector by two sets of effective
four--fermion couplings,
\beq
(G^s_{\rm NP})_{\alpha\beta}\,, \qquad (G^d_{\rm NP})_{\alpha\beta}\,,
\eeq
where $\alpha$ is the charged lepton index and $\beta$ is the flavor
of the neutrino in the weak interaction basis. At tree level, the
$SU(2)_L$ gauge symmetry implies that in the SM the four--fermion
couplings are proportional to $G_F\delta_{\alpha\beta}$. New
interactions, however, allow for non-diagonal and non-universal
couplings.

Phenomenological constraints imply that the new interactions are suppressed
with respect to the weak interactions. It is thus convenient to define small
dimensionless quantities in the following way:
\beq
\label{defeps}
\epsilon^p_{\alpha\beta}\equiv {(G_{\rm NP}^p)_{\alpha\beta}\over
\sqrt{|G_F+(G_{\rm NP}^p)_{\alpha\alpha}|^2+
\sum_{\gamma\neq\alpha} |(G_{\rm NP}^p)_{\alpha\gamma}|^2}}\,\qquad p=s,d.
\eeq
We denote by $|\nu_{\alpha}^{s}\rangle$ the neutrino states that is
produced at the source, and by $|\nu_{\alpha}^{d}\rangle$ the
neutrino state that is detected
\begin{equation}\label{nu-def}
|\nu_{\alpha}^{p}\rangle= 
\frac{G_{F}\delta_{\alpha\beta}+(G^p_{\rm NP})_{\alpha\beta}}{\sqrt{|G_F+(G_{\rm NP}^p)_{\alpha\alpha}|^2+\sum_{\gamma\neq\alpha}
|(G_{\rm NP}^p)_{\alpha\gamma}|^2}}|\nu_{\beta}\rangle\qquad p=s,d\,.
\end{equation}
At the leading order, we have
\beq
\label{nusd}
\epsilon^{p}_{\alpha\beta}=\frac{(G_{\rm NP}^p)_{\alpha\beta}}{G_{F}}\,,\qquad
|\nu_\alpha^p\rangle = |\nu_\alpha\rangle + \epsilon^p_{\alpha\beta}|\nu_\beta\rangle\,.
\eeq
The expression for the non-orthogonality parameter $\varepsilon_{\alpha\beta}$ at the
leading order is given by
\beq
\varepsilon_{\alpha\beta} \equiv
\vev{\nu_\alpha^s|\nu_\beta^d}=
\left\{
\begin{array}{lr}
1 + {\cal O}(\epsilon^2) & \quad \alpha=\beta\\
{\epsilon}^{s\,*}_{\alpha\beta}+\epsilon^d_{\beta\alpha} +{\cal O}(\epsilon^2)
& \quad
\alpha\neq \beta
\end{array}
\right.\,,
\eeq
where by ${\cal O}(\epsilon^2)$ we refer to effects that are quadratic in
$\epsilon^s$ or $\epsilon^d$. Note that in the SM the non-orthogonality
parameter vanishes, $\varepsilon_{\alpha\neq\beta}=0$.

For simplicity, we consider now a two generation model where the production
process is associated with a muon and the detection with a tau.
We calculate the following expression for the transition probability
\beq
P_{\mu\tau}=|\langle\nu_\tau^d|\nu_\mu^s(t)\rangle|^2,
\eeq
where $\nu_\mu^s(t)$ is the time-evolved state that was purely $\nu_\mu^s$
at time $t=0$. Using an explicit parameterization of the neutrino mixing matrix
\beq
U=\pcmatrix{\cos\theta & \sin\theta \cr -\sin\theta & \cos\theta}
\eeq
and keeping terms up to leading order in $\varepsilon$ we get
\beq
\label{nsi-osc}
P_{\mu\tau} =
\sin^2 x\left\{\sin^22\theta +
\Re(\epsilon^{d}_{\tau\mu} - \epsilon^{s}_{\mu\tau})
\sin
4\theta\right\}
+ \sin 2 x \;
\Im(\varepsilon_{\mu\tau})\;
\sin2\theta\,,
\eeq
where
\beq
\Delta m^2_{ij}\equiv m_i^2-m_j^2,\qquad  
x_{ij}\equiv{\Delta_{ij} L\over 4E},\qquad 
x=x_{12}\,,
\eeq
$m_i$ are the neutrino masses, $E$ is the neutrino
energy and $L$ is the distance between the source and the detector.

A few remarks are in order regarding Eq.~(\ref{nsi-osc}):
\begin{itemize}
\item
We keep only terms up to $O(\varepsilon)$. This is the reason why
there is no effect at $x=0$.
\item
The different $x$ dependence of the various terms is important as
it can be used to distinguish them experimentally.
\item
The interference term can be very important when $x \gg \varepsilon$.
\item
The interference term depends on the imaginary part of the NSIs, that
is, it requires CP violation.
\item
NSIs also affect the term proportional to $\sin^2 x$. Yet, within one
experiment this change is absorbed into the definition of $\theta$ and
cannot be distinguished experimentally.
\item
In many cases the NSIs are closely related to lepton flavor violating
charged lepton decays. However, they have a different dependence on
$\varepsilon$. Neutrino oscillation experiments are linear in $\varepsilon$
(if $x\gg\varepsilon$) while decays like $\tau\to\mu e^+ e^-$
are quadratic. This makes neutrino oscillation experiments competitive
in sensitivity.
\item
With three generations the result is more complicated and can be found
in \cite{GonzalezGarcia:2001mp}.
\end{itemize}

Before concluding this section, we remark on the case of heavy neutrinos.
For instance, consider the case of $k$ heavy singlet neutrinos. The mixing
matrix is $3 \times(3+k)$ and the $3\times 3$ mixing matrix for the light
neutrinos is not unitary anymore. In this case we have \cite{Antusch:2006vwa}
\beq
\varepsilon_{\alpha\beta} = \sum_h U_{h\alpha}U^*_{h\beta}~.
\eeq
The point is that using neutrino oscillation experiments we can
measure $\varepsilon$, and claim detection of some new physics, but we
can not disentangle the underlying mechanism which generates it.

\section{One loop NSI}
\label{Oneloop}

In this section, we extend the NSI formalism of
Refs.~\cite{Grossman:1995wx,GonzalezGarcia:2001mp} to include one-loop
effects. In particular, we consider universal NSIs from correction to
the tree level $W$ interaction and non-universal effects due to box
diagrams and scalar charged currents.

\subsection{Correction to the $W$ exchange amplitude}

At tree level, gauge invariance guarantees universality of the $W$
interactions. This universality is kept to all orders for an exact
symmetry. For a broken symmetry, however, universality is lost beyond
tree level. In the following, we show that one-loop effects make the
$W$ couplings and the kinetic terms of the fermions non-universal.
In particular, we explain why, in general, in the basis where the
kinetic term are canonical, the $W$ interactions are not flavor diagonal.

In the following, we neglect neutrino masses since they give subleading
effects to the NSIs, as we will discuss later.
Similarly, we do not consider other possible non-universal
flavor-diagonal NSIs.  They can be there but they are assumed to be
small and, to leading order, we can just add them to the effect we are
considering here.

On general grounds, one-particle irreducible one-loop effects include
the self energy diagrams for the charged leptons and the neutrinos,
and the corrections to the $W$ vertex as well. These one-loop diagrams
modify the kinetic and mass terms for fermions and the $W$ vertex by
factors $Z_L^\nu$, $Z_{L,R}^\ell$, and $\eta^{\ell}_{m}$ defined as
follows:
\begin{eqnarray}
{\cal L}_{\rm eff}
&=&\overline{\ell}_{jL} \left( Z_{L}^{\ell} \right)^{ji}
i\!\not\!\partial~\ell_{iL} +
\overline{\ell}_{jR}\left( Z^{\ell}_{R} \right)^{ji}
i\not\!\partial ~\ell_{iR} +
\overline{\nu}_{jL}\left( Z^{\nu}_{L}\right)^{ji}
i\!\not\!\partial~\nu_{iL}
\nonumber\\
&-&\overline{\ell}_{jR}\left(m^{\circ}_{\ell}+\eta^{\ell}_{m}\right)^{ji}
\ell_{iL}
-\overline{\ell}_{jL}\left(m^{\circ\,\dagger}_{\ell}+\eta^{\ell\dagger}_{m}\right)^{ji}
\ell_{iR}
\nonumber\\
&-&\frac{g}{\sqrt2}
W_\mu^-
\overline{\ell}_{jL}\gamma^\mu
\left(Z^{W}_{L}\right)^{ji}\nu_{iL} +{\rm H.c.}
\qquad\qquad\qquad i,j=1,2,3\,,
\end{eqnarray}
where the symbol ``$\circ$'' refers to ``bare'' or unrenormalized mass
matrices and $Z^a$ are matrices defined as
\beq
\left(Z^{a}\right)_{ij}= \delta_{ij} + \left(\eta^{a}\right)_{ij}
\qquad a=\nu,\ell,W\,.
\eeq
%
Note that Hermiticity of the effective Lagrangian ensures that
$(\eta^{\ell,\nu})^\dagger=\eta^{\ell,\nu}$ while $\eta^W$ may not be Hermitian.
While at tree level $\eta^a=0$, lepton flavor violating loop corrections introduce
non-zero $\eta$'s. In general, the off-diagonal elements of $\eta^a$ are finite,
while the diagonal terms diverge. Of course, the physics is finite and we discuss
how the divergences cancel below. We define
\beq
\hat{Z}^{\ell,\nu}_L = L^\dagger_{\ell,\nu} Z^{\ell,\nu}_L L_{\ell,\nu}~,
\qquad\qquad
\hat{Z}^{\ell}_R = R^\dagger_{\ell} Z^{\ell}_{R} R_{\ell}~,
\eeq
such that $\hat{Z}^a_{L,R}$ are diagonal matrices of positive
elements, and $L_\ell$ and $R_\ell$ are unitary matrices.  We rescale
the lepton fields to make their kinetic terms canonical
\begin{eqnarray}
\nu_{L}
\rightarrow
L_{\nu}\left(\hat{Z}^\nu_L\right)^{-\frac{1}{2}}~\nu_{L},
\qquad
\ell_{L}
\rightarrow
L_{\ell}\left(\hat{Z}^\ell_L\right)^{-\frac{1}{2}}~\ell_{L},
\qquad
\ell_{R}
\rightarrow
R_{\ell}\left(\hat{Z}^{\ell}_{R}\right)^{-\frac{1}{2}}~
\ell_{R}\,.
\label{eqn:rescaling}
\end{eqnarray}
where $(\hat{Z}^a)^{-1/2}$ is shorthand for the diagonal matrix of
element $(\hat{Z}^a)^{-1/2}_{ii}$.  The charged lepton mass terms
become
\begin{eqnarray}
&{\cal L}_{\rm mass}=
-\overline{\ell}_{jR}
\left(\hat{Z}^{\ell}_{R}\right)^{-\frac{1}{2}}R^{\dagger}_{\ell}
\bigg(m^{\circ}_\ell + \eta^{\ell}_{m} \bigg)
L_{\ell}\left(\hat{Z}^{\ell}_{L}\right)^{-\frac{1}{2}}
\ell_{iL}+\mbox{h.c.}
\label{eqn:effmassterm}
\end{eqnarray}
The mass terms (\ref{eqn:effmassterm}) can be diagonalized by two
independent rotations
\begin{eqnarray}
\ell_{L}\rightarrow L_m \ell_L,
\phantom{aaa}
\ell_{R}\rightarrow R_m \ell_R\,,
\label{eqn:rotations}
\end{eqnarray}
where $L_m$ and $R_m$ are unitary matrices. We obtain
\be
\left[R_m^\dagger\left(\hat{Z}^{\ell}_{R}\right)^{-\frac{1}{2}}R^{\dagger}_{\ell}
\bigg(m^{\circ}_\ell + \eta^{\ell}_{m} \bigg)
L_{\ell}\left(\hat{Z}^{\ell}_{L}\right)^{-\frac{1}{2}}L_{m}\right]^{ij}=
(m_{\ell})_{ij}\,,
\label{mass_corr}
\ee
where $m_{\ell}$ is diagonal.
After performing the rescaling (\ref{eqn:rescaling}) and the field rotation
(\ref{eqn:rotations}), the kinetic terms are canonical and the
charged lepton mass matrix is diagonal, whereas the interaction terms
are not diagonal. For later convenience, we rotate the neutrino fields
as $\nu_{L}\to L_m \nu_L$, in order to keep them as much aligned as
possible with their charged partners.  Note that this choice is allowed
because when we study NSIs we can neglect neutrino masses. As a result,
the $W$-boson vertices become
\beq
{\cal L}_{int}=
-\frac{g_2}{\sqrt{2}}W^{-}_\mu 
\overline{\ell}_L\gamma^\mu Z
\nu_L
-\frac{g_2}{\sqrt{2}}W^{+}_\mu 
\overline{\nu}_{L} \gamma^\mu Z^\dagger 
\ell_L,
\eeq
where 
\beq
Z =
L_{m}^{\dagger}
\left(\hat{Z}^{\ell}_L\right)^{-\frac{1}{2}}L^{\dagger}_{\ell}~
Z^W_L~
L_{\nu}\left(\hat{Z}^{\nu}_L\right)^{-\frac{1}{2}}~
L_{m}\,,
\label{eq:Wln_coupling_tot}
\eeq
Let us stress that eq.~(\ref{eq:Wln_coupling_tot}) is valid to all orders in
perturbation theory.
Finally, the relation to the physical parameter $\varepsilon_{\alpha\beta}^W$
can be derived from eq.~(\ref{nu-def}) and is given by
\beq \label{eq:Z-phys}
\varepsilon_{\alpha\beta}^W=\langle \nu^s_\alpha|\nu^d_\beta\rangle=
\frac{\left(ZZ^\dagger\right)_{\beta\alpha}}{\sqrt{(ZZ^\dagger)_{\alpha\alpha}
(ZZ^\dagger)_{\beta\beta}}}\,.
\eeq
This formula has a simple interpretation.
$Z_{\alpha\beta}/\sqrt{(ZZ^\dagger)_{\alpha\alpha}}~|\nu_{\beta}\rangle$
is the normalized state produced at the source, while its conjugate is
the one detected.

We proceed to find the leading order expressions for $Z_{\alpha\beta}$
and $\varepsilon^{W}_{\alpha\beta}$. That is, we will work to one loop
level. In this case, we can identify the off-diagonal terms of $Z$ with
the NSIs at the source and the detector, and therefore we find
\beq
Z_{\alpha\beta}=\epsilon^{s\,*}_{\alpha\beta}=\epsilon^{d\,*}_{\alpha\beta}\,,\qquad
\alpha\ne \beta.
\eeq
For the physical parameters we then get
\beq
\varepsilon^{W}_{\alpha\beta}=
Z_{\alpha\beta} + Z^{*}_{\beta\alpha}\,,
\qquad\qquad
\alpha\ne \beta\,.
\label{eq:Z_vs_epsilon}
\eeq
The same result can be obtained directly from eq.~(\ref{eq:Z-phys}). Note
that $ZZ^\dagger=\delta_{\alpha\beta}$ when $Z$ is unitary. When the
deviation from unitarity is small, $ZZ^\dagger=1+\varepsilon^W$, we
recover the previous result. See also eq.~(\ref{amplitude-formula}) below.

At one loop, the transformations for the lepton fields of
eq.~(\ref{eqn:rescaling}) read
\begin{eqnarray}
\nu_{L}
\rightarrow
\left(1 - {1\over2}\eta_{L}^{\nu}\right)
\nu_{L},
\qquad
\ell_{L}
\rightarrow
\left(1 - {1\over2}\eta_{L}^{\ell}\right)
\ell_{L},
\qquad
\ell_{R} 
\rightarrow
\left(1 - {1\over2}\eta_{R}^{\ell}\right) \ell_{R}\,.
\label{eqn:rescaling_1loop}
\end{eqnarray}
Similarly, to one loop accuracy, the transformations of
eq.~(\ref{eqn:rotations}) read
\begin{eqnarray}
\ell_L \rightarrow \left( 1 + \delta L_m \right)\ell_L,
\phantom{aaa}
\ell_R \rightarrow \left( 1 + \delta R_m \right)\ell_R\,.
\label{eqn:rotations_apx}
\end{eqnarray}
The unitarity of $L_m$ and $R_m$ implies that
\beq
\delta L^\dagger_{m}=-\delta L_{m}\,,\qquad \delta R^\dagger_{m}=-\delta R_{m}\,.
\label{eqn:unitarity_condition}
\eeq
In this approximation, $Z$ is given by
\beq
Z = 1 + \eta^W_L - \frac{\eta^{\ell}_{L}+\eta^{\nu}_{L}}{2}\,,
\label{eq:Wln_coupling_tot_1loop}
\eeq
where $\eta^\nu$, $\eta^\ell$, and $\eta^W$ have to be evaluated at
one loop accuracy.  Eq.~(\ref{eq:Wln_coupling_tot_1loop}) shows that,
at this level, the dependence of $Z$ on $L_m$ drops completely out
(its dependence would be reintroduced at two loop). Then we learn
that at leading order the non-orthogonality parameter between the
source and detector neutrinos is given by (\ref{eq:Z_vs_epsilon})
and reads
\begin{equation}
\label{main}
\varepsilon^{W}_{\alpha\beta}=\epsilon^{s\,*}_{\alpha\beta}+\epsilon^{d}_{\beta\alpha}=
\left(\eta^{W\dagger}_L + \eta^{W}_L - \eta^{\ell}_L - \eta^\nu_L\right)_{\alpha\beta}\,,
\end{equation}
where we used the fact that $\eta^\ell$ and $\eta^\nu$ are Hermitian.

Eqs.~(\ref{eq:Wln_coupling_tot}), (\ref{eq:Z_vs_epsilon}), and
(\ref{main}) are the main results of this section. A few remarks are in
order when inspecting them:
\begin{enumerate}
\item
For any given model, eqs.~(\ref{eq:Wln_coupling_tot}),
(\ref{eq:Z_vs_epsilon}) show how to extract the physical NSI effects
out of the calculations of the vertex corrections and the self
energies encoded in the rotation mass matrices $L_\nu$, $L_\ell$ and
$L_m$.  However, eq.~(\ref{main}) demonstrates that, at one loop
accuracy, all we need to do is to calculate $\eta^\nu$, $\eta^\ell$,
and $\eta^W$, since $L_m$ starts contributing only at two loop.
\item
We note that $\varepsilon^{W}_{\alpha\beta}$ is finite. While each of
the diagonal terms in $\eta^a$ may diverge, the combination is finite
because of the $SU(2)_L$ gauge symmetry.  This can be seen by the
fact that the UV properties are insensitive to EWSB. Thus, the divergent
part of $\eta^a$ is real, flavor universal and independent of
$a$. Therefore, the divergent part of
$(\eta^{W\,\dagger}_L+\eta^{W}_L-\eta^{\ell}_L-\eta^\nu_L)_{\alpha\alpha}$,
as well as its renormalization scale dependence, cancel. In contrast,
the flavor off-diagonal elements of $\eta^{W}_{L}$ and
$\eta^{\nu,\ell}_{L}$ are singularly finite and scale independent, as
a result of the GIM mechanism. We stress also that when the
electroweak symmetry is restored, $v_{EW}\rightarrow 0$, then the
finite parts are flavor universal and
$\epsilon^p_{\alpha\beta}\rightarrow 0$. Both results are illustrated
in a concrete example below where the $\eta^a$ are explicitly
calculated.
\item
Considering the CP-conjugated process, we obtain
\beq
\mathcal{A}_{NSI}^{CP}= \varepsilon^{W}_{\beta\alpha}= \varepsilon^{W\,*}_{\alpha\beta}\,.
\eeq
So, CP is violated when the $\eta^a$ have non trivial imaginary parts.
\item
Eq.~(\ref{main}) admits an interpretation in terms of the scattering
amplitudes $\mathcal{M}_{\alpha\beta}$. The idea is to think about
neutrinos as invisible intermediate states, and sum over all of them
in the propagation process from the source to the
detector. Considering a source neutrino associated with a charged
lepton $\alpha$ and a detection that is done by a charged lepton
$\beta$, the scattering amplitude is given by
\beq
\label{amplitude-formula}
\mathcal{M}_{\alpha\beta} \propto (Z Z^\dagger)_{\beta\alpha}=
\delta_{\beta\alpha}+
\left(\eta_{L}^{W}+\eta^{W\dagger}_{L}-
\eta_{L}^{\nu}-\eta_{L}^{\ell}\right)_{\beta\alpha}=
\delta_{\alpha\beta}+ \varepsilon^{W}_{\beta\alpha}\,.
\eeq
where in the second step we expanded in $\eta_{L}^{\nu,l,W}$.
Thus, we learn that the non-universal part of the amplitude
is just $\varepsilon^{W}_{\beta\alpha}$.
\end{enumerate}

\subsection{Non-universal effects}

So far, we have considered only NSIs induced by the universal
corrections to the $W$ vertex and self energy diagrams. These effects
are independent of the production or detection processes as long as
these processes are mediated by $W$ exchange.  We now move to discuss
other loop effects that are not universal, that is, that may be
different for different production or detection processes.

As mentioned before, the universal effects are $SU(2)$ breaking effects
and therefore suppressed like $M^2_W/M^2_{NP}$.  In the effective
field theory language, this would correspond to the effects induced,
after the EWSB, by gauge invariant dimension-six operators like
$(\bar{L}_L \tau^a\gamma^\mu L_L)(H^\dagger \tau^a D_\mu H)$, where
$L_L$ ($H$) stands for the lepton (Higgs) doublet, $\tau^a$ are either
the identity or the $SU(2)$ generators and $D_\mu$ is the covariant
derivative.

It is then clear that contributions arising from dimension six
four-fermion operators, also suppressed by $M^2_W/M^2_{NP}$, must be
consistently included along with the $W$-penguin
contributions. Therefore, for any given New Physics (NP) model we can write the
following expression for the physical $\varepsilon_{\alpha\beta}$
parameter
\begin{equation}
\label{main_w_box}
\varepsilon_{\alpha\beta} =
\varepsilon^{W}_{\alpha\beta} + \left(\epsilon^{s\,*}_{\alpha\beta}\right)^{dim-6}+
\left(\epsilon^{d}_{\beta\alpha}\right)^{dim-6}\,.
\end{equation}
where $\varepsilon^{W}_{\alpha\beta}$ has been already defined in
eq.~(\ref{main}) and we have assumed both matter effects and higher
order operators (with dim $>6$) to be negligible.

In this work we are interested in NSIs with the same Dirac structure as the
SM interactions. The reason is that they maximally exploit the interference
between SM and NP amplitudes. In particular, focusing on realistic production
and detection processes like $\mu\to e\nu_e\bar{\nu}_\alpha$ and $P\to\mu\nu_\alpha$
(with $P=\pi,K$), the relevant dimension six operators are the following
%
\begin{eqnarray}
&&\frac{4 G_F}{\sqrt 2}
\left(\delta_{\mu\alpha} + (\epsilon^{s}_{\mu\alpha})^{dim-6}\right)
\left(\overline{\nu}_{\alpha}\gamma^\lambda P_L\mu \right)
\left(\overline{e}\gamma_\lambda P_L \nu_e \right)\,,
\label{dim6_muon}
\\
&&\frac{4 G_F}{\sqrt 2}\left(\delta_{\mu\alpha}+(\epsilon^{d}_{\mu\alpha})^{dim-6}\right)
\left(\overline{u}\gamma^\lambda P_L d \right)
\left(\overline{\mu}\gamma_\lambda P_L \nu_\alpha \right)\,,
\label{dim6_plnu_detercor}
\end{eqnarray}
where $(\epsilon^{s,d}_{\mu\alpha})^{dim-6}$ stand for the loop-induced corrections.

As we will discuss in the next section, in the context of
supersymmetry $\epsilon^{\rm dim-6}_{\mu\alpha}$ might be induced
either by means of gaugino/sfermion boxes or through the tree level
exchange of charged Higgs with loop induced flavor changing couplings
$H\ell\nu$.

\subsection{Matter effects}

So far, we have discussed only NSIs induced by the charged currents.
However, we would like to emphasize here that there exist also NSIs in
matter via neutral currents, even for negligible SM matter effects.

Indeed, starting from the SM neutral current interactions
\begin{eqnarray}
{\cal L}_{\rm eff}=
-\frac{g}{2 c_W}
Z_\mu \overline{\nu}_{jL}\gamma^\mu
\left(Z^{Z}_{L}\right)^{ji}\nu_{iL}
+{\rm H.c.}~~i,j=1,2,3\,,
\end{eqnarray}
where hereafter $c_W=\cos\theta_W$ and $s_W=\sin\theta_W$, after performing the
rescaling (\ref{eqn:rescaling}) and the field rotation (\ref{eqn:rotations}),
the $Z$-boson vertex with neutrinos is given by
\beq
Z =
L_{m}^{\dagger}
\left(\hat{Z}^{\nu}_L\right)^{-\frac{1}{2}}L^{\dagger}_{\nu}~
Z^Z_L~
L_{\nu}\left(\hat{Z}^{\nu}_L\right)^{-\frac{1}{2}}~
L_{m}\,,
\label{eq:Znn_coupling_tot}
\eeq
where, again, we rotated the neutrino fields as $\nu_{L}\to L_m \nu_L$.
At one loop accuracy, eq.~(\ref{eq:Znn_coupling_tot}) becomes
\beq
Z = 1 + \eta^Z_L - \eta^{\nu}_{L}\,.
\label{eq:Znn_coupling_tot_1loop}
\eeq
We stress that also NSIs induced by the $Z$-penguin are $SU(2)$
breaking effects and all the considerations made for $W$-penguin
induced NSIs apply also here.  Therefore, besides $Z$-penguin effects,
we have to include dimension six four-fermion operators. The relevant
neutral interactions are
\begin{eqnarray}
&& \sqrt 2 G_F
\bigg[
\varepsilon^{Z}_{\mu\alpha}\left( I^{f}_{3L} - 2 Q_f s^2_W \right) +
(\varepsilon^{m,f}_{\mu\alpha})^{dim-6}
\bigg]
\left(\overline{\nu}_{\alpha}\gamma^\lambda P_L\nu_\mu \right)
\left(\overline{f}\gamma_\lambda f \right)\,,
\label{dim6_matter}
\end{eqnarray}
where $f$ stands for the fermions in the matter. For normal matter, $f$ could be
electrons, protons and neutrons $f = e,p,n$, $Q_f$ is the electric charge of $f$ 
and $I^{f}_{3L}$ is the third component of weak isospin of the left-chiral projection 
of $f$.

Since we can identify $\varepsilon^{Z}_{\mu\alpha}={Z}_{\mu\alpha}$, the total
physical parameter in the matter $\varepsilon^{m,f}_{\mu\tau}$ reads
\beq
\varepsilon^{m,f}_{\mu\tau}=
\left( I^{f}_{3L} - 2 Q_f s^2_W \right)
\left( \eta^{Z}_{L} - \eta^{\nu}_{L} \right)_{\mu\tau} +
(\varepsilon^{m,f}_{\mu\tau})^{dim-6}
\,.
\eeq
When matter effects are included, the transition probability $P_{\mu\tau}$ of
Eq.~(\ref{nsi-osc}) becomes
\bea
\label{nsi-osc-matter}
P_{\mu\tau} &\simeq&
x~\sin2\theta \bigg[ x~\sin2\theta - 2L \sum_f A^f~\Re(\varepsilon^{m,f}_{\mu\tau}) \bigg]
\nonumber\\
&+&
x^2~\Re(\epsilon^{d}_{\tau\mu} - \epsilon^{s}_{\mu\tau})\sin 4\theta
+ 2 x~\Im(\varepsilon_{\mu\tau})\sin2\theta\,,
\eea
where $A^f = \sqrt{2}G_F n_f$ and we have assumed $x\ll 1$ and constant fermion densities $n_f$.

A close look to Eq.~(\ref{nsi-osc-matter}) shows that the interference term
between the SM and non-SM matter effects ($\varepsilon^{m,f}_{\mu\tau}$) depends
only on the real part of $\varepsilon^{m,f}_{\mu\tau}$. By contrast, for NSIs
at the production or detection processes ($\varepsilon^{s,d}_{\mu\tau}$),
the interference term depends only on the imaginary part of the NSIs.

The situation changes when considering the transition probabilities $P_{e\mu}$
and $P_{e\tau}$ which involve the electron neutrino. In these cases, there are
interference terms, driven by charged current SM-effects, which are also
sensitive to the real parts of $\varepsilon^{s,d}$~\cite{GonzalezGarcia:2001mp}.


\subsection{Scalar charged current}

Many UV completions of the SM contain an extended Higgs sector, for
example, the MSSM. On general grounds, the presence of at least two
Higgs doublets leads to a misalignment between the fermion mass
matrices and the Yukawa couplings. As a result, Higgs mediated FCNC
processes are induced already at tree level resulting in large effects,
unless a flavor protection mechanism is at work. Indeed, the Natural
Flavor Conservation (NFC) hypothesis was introduced to deal with this
flavor problem. However, even if NFC holds at the tree level, this
hypothesis is spoiled by quantum corrections~\cite{Buras:2010mh}.
For instance, if NFC arises as a result of a continuous PQ symmetry,
the breaking at the quantum level of such a symmetry (as it is required
in order to prevent the appearance of a massless Goldstone boson) would
reintroduce FCNC effects~\cite{Buras:2010mh}.

This is the case of supersymmetry where the holomorphy of the superpotential
implies a type-II structure of the Higgs potential at the tree level.
Yet, the presence of a non-vanishing $\mu$-term (such that $\mu H_u H_d$)
induces, after SUSY breaking, non-holomorphic Yukawa couplings for fermions
(such as $\bar{Q}_L d_R H^{\dagger}_{u}$)~\cite{Hall:1993gn} and therefore
Higgs-mediated flavor violation is unavoidable~\cite{Hamzaoui:1998nu}.

Bearing in mind the above considerations, in the following we perform a model
independent analysis of NSIs arising from loop-induced scalar charged currents.

The charged Higgs $H^\pm$ couplings with leptons are described by the
following effective Lagrangian
\begin{eqnarray}
{\cal L}_{\rm eff}
=\overline{\nu}_{jL}
\left( y^{\circ}_{\ell} + \eta^H \right)^{ji}\ell_{iR} H^+
+{\rm H.c.}
\end{eqnarray}
where 
\beq
y^{\circ}_\ell = {g_2\over\sqrt2M_W}m^{\circ}_\ell\tan\beta
\eeq
and we recall that ``$\circ$'' refers to unrenormalized quantities.

The transformations of eqs.~(\ref{eqn:rescaling}), (\ref{eqn:rotations}) lead to
the following effective Lagrangian valid to all orders in perturbation theory
\begin{eqnarray}
{\cal L}_{\rm eff}
=\overline{\nu}_{jL}
\left[L_{m}^{\dagger}\left(\hat{Z}^{\nu}_{L}\right)^{-\frac{1}{2}}L^{\dagger}_{\nu}
\bigg( y^{\circ}_{\ell} + \eta^H \bigg)
R_{\ell}\left(\hat{Z}^{\ell}_{R}\right)^{-\frac{1}{2}} R_{m}\right]^{ji}
\ell_{iR} H^+
+{\rm H.c.}\,.
\label{eq:Hln}
\end{eqnarray}
In order to find the leading expansion for the above effective couplings,
we proceed as follows. We first rescale the lepton fields at one loop level,
see eq.~(\ref{eqn:rescaling}), so that the charged lepton mass terms become
\begin{eqnarray}
&&{\cal L}_{\rm mass}=
-\overline{\ell}_{jR}\left(m^{\circ}_\ell+\delta m_\ell\right)^{ji}\ell_{iL}+\mbox{h.c.}\,,
\label{eqn:effmassterm_1loop}
\end{eqnarray}
where
\begin{eqnarray}
\delta m_\ell
\equiv
\eta_{m}^{\ell}-
{1\over2}\eta_{R}^{\ell} m^{\circ}_{\ell}-
{1\over2}m^{\circ}_{\ell} \eta_{L}^{\ell}~.
\label{eqn:deltam}
\end{eqnarray}
Then, we make use of the one loop expansions for the matrices
$L_m$ and $R_m$ of eq.~(\ref{eqn:rotations_apx}) leading to
\be
\bigg[m^{\circ}_\ell + \delta m_\ell -\delta R_m m^{\circ}_\ell + m^{\circ}_\ell\delta L_m \bigg]^{ji} =
m_{\ell}^{ji}\,,
\label{mass_corr_apx}
\ee
where $m_{\ell}$ is diagonal and we have consistently retained only
the leading one-loop terms.
At this level, the unitarity condition of eq.~(\ref{eqn:unitarity_condition})
ensures that ${\rm Re}(\delta L^{jj}_{m})={\rm Re}(\delta R^{jj}_{m})=0$ and
the corrected mass eigenvalues are given by
\begin{eqnarray}
m_{\ell_j}= m^\circ_{\ell_j} + {\rm Re} \left(\delta m_\ell\right)_{jj}\,,
\label{eqn:masscorr}
\end{eqnarray}
while the condition of reality for the masses implies
\begin{eqnarray}
{\rm Im}~\delta R^{jj} - {\rm Im}~\delta L^{jj} =
\frac{{\rm Im} \left(\delta m_\ell\right)_{jj}}{m_{\ell_j}}\,.
\label{eqn:mass_real}
\end{eqnarray}
Finally, $\delta L_{m}$ and $\delta R_{m}$ are determined by
\begin{eqnarray}
\delta L^{ji}_{m}=
{m_{\ell_j} (\delta m_\ell)^{ji} + (\delta m_\ell^\dagger)^{ji} m_{\ell_i}
\over m_{\ell_i}^2- m_{\ell_j}^2} ~~~j\neq i\,,
\label{eqn:dL}
\end{eqnarray}
\begin{eqnarray}
\delta R^{ji}_{m}=
{m_{\ell_j}(\delta m_\ell^\dagger)^{ji} + (\delta m_\ell)^{ji} m_{\ell_i}
\over m_{\ell_i}^2 - m_{\ell_j}^2} ~~~j\neq i\,,
\label{eqn:dR}
\end{eqnarray}
where $m_{\ell_i}=(m_\ell)_{ii}$, that is, it is the $i$th eigenvalue
of $m_\ell$.
We are ready now to expand eq.~(\ref{eq:Hln}) up to one loop. By
making use of the eqs.~(\ref{eqn:deltam}), (\ref{eqn:masscorr}), and
(\ref{eqn:mass_real}), we obtain the following flavor conserving
couplings
\be
{\cal L}^{H^+}_{\rm eff}\!=\!
\overline{\nu}_{iL}
\left[
y_{\ell_i} \left( 1 + {1\over2}\eta_{L}^{\ell} - {1\over2}\eta_{L}^{\nu}
- {\eta_{m}^{\ell\dagger}\over m_{\ell_i}} \right)
+ \eta^H
\right]^{ii}
\!\!
\ell_{iR} H^+ +{\rm H.c.}
\label{Hln_LFC}
\ee
where 
\beq
y_{\ell_i} =\frac{g_2 m_{\ell_i}}{\sqrt{2}M_W}\tan\beta.
\eeq
Again, while each term in eq.~(\ref{Hln_LFC}) is in general divergent
and renormalization scale dependent, their sum is finite and scale
independent.

For the flavor violating charged Higgs couplings, we find the one loop expression
\be
{\cal L}^{H^+}_{\rm eff}=
\overline{\nu}_{jL}
\left[- y_{\ell_i}\delta L_{m} + y_{\ell_j}\delta R_{m} - {1\over2}y_{\ell_i}\eta_{L}^{\nu} - {1\over2}y_{\ell_j}\eta_{R}^{\ell} + \eta^H
\right]^{ji}
\!\!\!
\ell_{iR} H^+ +{\rm H.c.}\,,
\label{Hln_LFV}
\ee
and each term in eq.~(\ref{Hln_LFV}) is finite and scale independent
thanks to the GIM mechanism. Notice that, in contrast to the case of
NSIs at the $W$-boson vertex, we are now sensitive to the rotation
mass matrices $L_m$ and $R_m$ already at the one loop level. As we
will discuss in the next section, within the MSSM the rotations
$\delta L_m$ and $\delta R_m$ actually provide the dominant effects to
the flavor changing couplings.

Let us consider now the case of $j=3$, that is relevant for a tau neutrino
production. In such a case, the one loop expansions for $\delta
L_m$ and $\delta R_m$ of eqs.~(\ref{eqn:dL}) and (\ref{eqn:dR}) take
the form
\bea
\left(\delta R_m \right)^{3i}
&\simeq&
\left[ -{\eta_{m}^{\ell\dagger} \over m_{\tau}}
       + {1\over2}\eta_{R}^{\ell}
       -{m_\mu \over m_\tau} {\eta_{m}^{\ell\dagger} \over m_{\tau}}
       + {m_\mu \over m_\tau} \eta_{L}^{\ell}
\right]^{3i}\,,
\\
\left(\delta L_m \right)^{3i}
&\simeq&
\left[ -{\eta_{m}^{\ell} \over m_{\tau}}
       + {1\over2}\eta_{L}^{\ell}
\right]^{3i}\,.
\eea
Finally, inserting the above expressions for $\delta L_m$ and $\delta R_m$
in eq.~(\ref{Hln_LFV}), we find
\be
{\cal L}^{H^+}_{\rm eff}=
\overline{\nu}_{\tau L}~Z_{H}^{3i}~\ell_{iR} H^+ + {\rm H.c.}\,,
\label{Hln_LFC_final}
\ee
where
\be
Z_{H}^{3i} =
\left[~{1\over2} y_{\ell_i} \eta_{L}^{\ell} - {1\over2} y_{\ell_i} \eta_{L}^{\nu}
      -y_{\tau}{\eta_{m}^{\ell\dagger}\over m_{\tau}} + \eta^H~\right]^{3i}\,.
\!\!\!
\label{eq:ZH}
\ee
Let us remark that NSI effects driven by charged scalar currents are
expected to be particularly relevant for the neutrino production via
charged meson decays.  In fact, whenever the relevant Yukawa couplings
are proportional to the fermion masses, only processes like
$P\to\ell\nu$ (with $P=K,\pi$), which are helicity suppressed in the
SM, might receive large contributions. Other production processes like
$\mu\to e\nu\bar{\nu}$ or detection cross-sections are expected not to be
significantly affected by such charged scalar currents. As a result,
we now have $\epsilon^{s}_{\alpha\beta}\gg\epsilon^{d}_{\alpha\beta}$
and therefore
$\varepsilon_{\mu\tau}\approx\epsilon^{s\,*}_{\mu\tau}$. This is in
contrast to the case with dominant NSIs at the W-boson vertex where,
as we already discussed, it turns out that
$\epsilon^{s}_{\alpha\beta}=\epsilon^{d}_{\alpha\beta}$.

We can proceed now to establish the relation between
$\varepsilon_{\mu\tau}$ and $Z_{H}^{32}$ in the case where the
neutrino source is given by the process $\pi\to\mu\nu$. This decay is
mediated by tree level $W^\pm$ and $H^\pm$ exchanges. The relevant
effective Lagrangian describing this process is
\beq
\frac{4G_F}{\sqrt{2}}V_{ud}
(\,\overline{u}\gamma_{\mu}P_L d\,)(\,\overline{\mu}\gamma^{\mu}P_L\nu_\mu\,)+
V_{ud}\left(\frac{y_d Z^{32\,*}_H}{m^{2}_{H^\pm}}\right)
(\,\overline{u}P_R d\,)(\,\overline{\mu} P_L \nu_{\tau})\,,
\eeq
where $P_{R,L}=(1\pm \gamma_5)/2$ and $y_d$ is the down quark Yukawa coupling.
Since the $\pi$ meson is a pseudoscalar, its decay amplitude can be induced only
by the axial-vector part of the $W^\pm$ coupling and the pseudoscalar part of 
the $H^\pm$ coupling. Then, once we implement the PCAC relations
\beq
\langle0|\overline{u}\gamma_{\mu}\gamma_{5}d|\pi^{-}\rangle=
if_\pi p^{\mu}_\pi\,, \qquad
\langle0|\overline{u}\gamma_{5}d|\pi^{-}\rangle=
-if_\pi \frac{m^{2}_\pi}{m_{d}+m_u}\,,
\eeq
we get the amplitudes
\bea
\mathcal{M}^W_{\pi\to \mu\nu_\mu} &=&
\frac{G_F}{\sqrt{2}} V_{ud}f_\pi m_\mu~ \overline{\mu}(1-\gamma_{5})\nu_\mu~,
\\
\mathcal{M}^H_{\pi\to \mu\nu_\tau} &=&
- {V_{ud}f_\pi\over 4}
\bigg(\frac{y_d Z^{32\,*}_H}{m^{2}_{H^\pm}}\bigg)\frac{m^{2}_\pi}{m_d\!+\!m_u}~
\overline{\mu}(1-\gamma_{5})\nu_\tau
\,.
\eea
We observe that the SM amplitude depends on the lepton mass because of
the helicity suppression, in contrast to the charged Higgs amplitude
which does not suffer from this suppression. Yet, many NP models
predict the Yukawa couplings to be proportional to the fermion
masses. Effectively then, in such models both the $W$-boson and
charged Higgs amplitudes have the same lepton mass dependence.

Finally, recalling that the produced neutrino state is
$|\nu^{s}\rangle=|{\nu}_{\mu}\rangle+\epsilon^s_{\mu\tau}|{\nu}_{\tau}\rangle$,
we identify
\beq
\varepsilon^{\pi}_{\mu\tau}
\approx
\left(\epsilon^{s\,*}_{\mu\tau}\right)^{\pi} = -
\frac{\sqrt{2}}{4 G_F}
\bigg(\frac{m^{2}_\pi}{m_d+m_u}\bigg)
\bigg(\frac{y_d Z^{32}_H}{m_\mu m^{2}_{H^\pm}}\bigg)\,.
\label{eq:eps_H}
\eeq
In the case of $K\to\mu\nu$, the relevant parameter $\varepsilon^{K}_{\mu\tau}$
can be simply obtained from $\varepsilon^{\pi}_{\mu\tau}$ through the replacement
$(m_\pi,y_d,m_d)\to (m_K,y_s,m_s)$ and we get
\beq
\frac{\varepsilon^{\pi}_{\mu\tau}}{\varepsilon^{K}_{\mu\tau}}
\simeq
\frac{m^{2}_{\pi}}{m^{2}_{K}}~\frac{m_s+m_u}{m_d+m_u}~\frac{y_d}{y_s}
\sim \frac{1}{20}\,.
\label{eq:eps_H_pi_vs_k}
\eeq
It has yet to be seen which process $\pi\to\mu\nu$ or $K\to\mu\nu$ may
represent the best probe of this scenario when combining the more intense
neutrino flux obtainable from $\pi\to\mu\nu$ with the higher NP sensitivity
of $K\to\mu\nu$.

\section{One loop NSIs and Supersymmetry}
\label{nsi_susy}

In this section, we apply the model-independent formalism developed in
the previous section to the case of the R-parity conserving MSSM with 
new sources of LFV in the soft sector. We will analyse first loop-induced
NSIs from the $V-A$ charged current, passing then to NSIs from the charged 
scalar current induced by the heavy Higgs sector of the MSSM.

\subsection{$V-A$ charged current}

In the MSSM NSIs may be induced by the $V-A$ charged current through
$W$-penguin as well as box contributions. The former effect arises
only after the EWSB and therefore is suppressed by
$M^2_W/M^2_{NP}$. In particular, within the MSSM, there are three
possible sources of $SU(2)$ breaking: 
\begin{itemize}
\item[i)]
the $D$-terms, 
\item[ii)]
the left-right mixing terms, and 
\item[iii)]
the neutralino/chargino mixing terms. 
\end{itemize}
The latter comes from dimension six four-fermion operators and
is also suppressed by $M^2_W/M^2_{NP}$.

The full analytical calculation relevant for NSIs in the MSSM, upon
which our numerical analysis is based on, is reported in the appendix.
In the following, instead, we prefer to discuss general properties
within an illustrative toy model which is a particular limit of the
MSSM that greatly simplifies the calculation but still retains the
most relevant features.
We use standard MSSM notation (for a review see for example,
Ref.~\cite{Martin:1997ns}) and we consider only the lepton sector.

In our toy model, we decouple the Higgsinos, $\tilde{H}$, by taking a
large $\mu$-parameter and we decouple the Winos, $\tilde W$, by taking
their soft mass term, $M_{2}$, to be very large.  We also assume
left-right mixing between the sleptons to be negligible. Finally, we
take the EW vacuum expectation value, $v_{EW}$, small compared to the
soft mass term of the Bino, $M_{1}$, and the soft term for the
left-handed sleptons, $M_{L}$. That is, we consider a model where
\beq
\frac{A}{M_{L}}~, \frac{m_\tau\mu\tan\beta}{M^2_L} \ll 1~,
\qquad
v_{EW}\ll M_{1}\sim M_{L}\ll \mu\sim M_{2}\,.
\eeq
where $A$ stands for the trilinear soft terms. In this model there is
only one neutralino, the Bino. In practice, the model looks
supersymmetric only with respect to $U(1)_Y$. The EWSB can be treated
as a perturbation.
Note that, in our toy model, only the $SU(2)$ breaking source i) is at
work.  Later on, we will also discuss the impact on NSIs of sources
ii) and iii), which are expected in a more realistic model.

The relevant interactions are the lepton-Bino-slepton vertex and the slepton-slepton-$W$
vertex, that are given by
\beqa
\mathcal{L}_{NSI}= -\sqrt{2}ig^{\prime} q_{Y}\bar{\chi}^0
\left(\tilde{\nu}^{*}_k U^{ki}_{\tilde{\nu}}\nu_{i} +
\tilde{\ell}^{*}_k  U^{ki}_{\tilde{\ell}} \ell_{i}\right) -\sqrt{2}ig
\left(W^\mu \partial_{\mu}\tilde{\nu}_{k}\tilde{\ell}^*_k\right)+h.c.
\eeqa
where $q_{Y}=-1/2$ is the hypercharge of the left-handed leptons.
$U^{ki}_{\tilde{\nu}}$ ($U^{ki}_{\tilde{\ell}}$) is the unitary matrix
that diagonalizes the sneutrino (charged slepton) mass matrix. In our
model, the soft terms are $SU(2)_L\times U(1)_{Y}$ symmetric and, since
there are no left-right mixings, $U_{\tilde{\nu}}=U_{\tilde{\ell}}$.
Thus, from this point on we use $U_{\tilde{\ell}}$ for both terms.

\begin{figure}[tb]
\begin{center}
\includegraphics[scale=1.5]{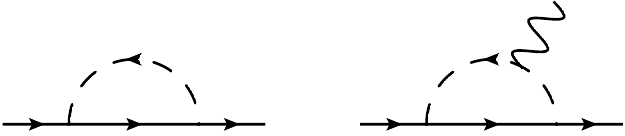}
\caption{1-loop contributions to the lepton self-energies (left) and to the vertex (right).
The virtual particles running in the loop are sleptons and a neutralino.}%
\label{1loop-W-NSI}%
\end{center}
\end{figure}

According to Eq.~(\ref{main}), all we need to calculate are the loops in
Fig.~\ref{1loop-W-NSI} and extract $\eta^\nu$, $\eta^\ell$ and $\eta^W$.
In our calculations we use a naive UV cutoff, that is, we perform the
$k^2$ integral up to $\Lambda^2$. We further introduce an unphysical
mass parameter $\mu$ to make the arguments of all logarithms dimensionless.
Effectively, this is equivalent to introducing a renormalization scale.
As a check on our calculation we see that the final physical results,
that is, $\varepsilon$, is independent of these two unphysical parameters.

We first calculate the wave function corrections. The only diagrams contributing
to $\eta^\nu$ at one loop are through the exchange of the Bino and sneutrinos.
The result is
\beq \label{etanu} 
\eta^{\nu}_{ji}= {g^{\prime}}^2q_{Y}^2\sum_{k}
U^{kj\,*}_{\tilde{\ell}} U^{ki}_{\tilde{\ell}}~\mathcal{I}^\nu_{k}\,,
\eeq
where
\beqa
\mathcal{I}^\nu_k&=& \frac{1}{16\pi^2}
\left\{\log\frac{\Lambda^2}{\mu^2}+F_k^\nu+
\mathcal{O}\left(\frac{1}{\Lambda^2}\right)\right\}, \\
F_k^\nu &=&\left[
-\log\left(\frac{m^2_{\tilde{\nu}_k}}{\mu^2}\right)+
\frac{m^4_{\chi^0}}{(m^2_{\tilde{\nu}_k}-m^2_{\chi^0})^2}
\log\left(\frac{m^2_{\tilde{\nu}_k}}{m^2_{\chi^0}}\right)
-\frac{m^2_{\tilde{\nu}_k}+m^2_{\chi^0}}{2(m^2_{\tilde{\nu}_k}-m^2_{\chi^0})}
\right]\,.
\eeqa

The calculation of the wave function correction to the left-handed
charged leptons $\eta^{\ell}_{ji}$ proceeds in a completely analogous
way, the only difference being that now we have sleptons, instead of
sneutrinos, running in the loop.  Therefore, $\eta^{\ell}_{ji}$ reads
\beq \label{etaell} 
\eta^{\ell}_{ji}={g^{\prime}}^2q_{Y}^2\sum_{k}
U^{kj\,*}_{\tilde{\ell}} U^{ki}_{\tilde{\ell}}~\mathcal{I}^\ell_{k}\,,
\eeq
where $\mathcal{I}^\ell_{k}$ is simply obtained from $\mathcal{I}^\nu_{k}$
by the replacement $m^2_{\tilde{\nu}_k}\to m^2_{\tilde{\ell}_k}$.

Next, we calculate the one loop correction to the $W$ vertex, $\eta^W$.
The result is
\beq \label{etaW}
\eta^W_{ij}=
{g^{\prime}}^2q_{Y}^2 \sum_{k}
U^{kj\,*}_{\tilde{\ell}} U^{ki}_{\tilde{\ell}}~\mathcal{I}^W_{k},
\eeq
where
\beqa
\mathcal{I}^W_{k}&=& \frac{1}{16\pi^2}
\left\{\log\frac{\Lambda^2}{\mu^2}+F_{k}^W+
\mathcal{O}\left(\frac{1}{\Lambda^2}\right)\right\}\,,\\
\nonumber
F_{k}^W &=&\frac{1}{(m^2_{\tilde{\nu}_k}-m^2_{\tilde{\ell}_k})(m^2_{\tilde{\nu}_k}-m^2_{\chi^0})(m^2_{\tilde{\ell}_k}-m^2_{\chi^0})}\left[m^4_{\tilde{\nu}_k}(m^2_{\chi^0}-m^2_{\tilde{\ell}_k})\log\left(\frac{m^2_{\tilde{\nu}_k}}{\mu^2}\right) \right.\\
&&\left. +
m^4_{\tilde{\ell}_k}(m^2_{\tilde{\nu}_k}-m^2_{\chi^0})\log\left(\frac{m^2_{\tilde{\ell}_k}}{\mu^2}\right)+
m^4_{\chi^0}(m^2_{\tilde{\ell}_k}-m^2_{\tilde{\nu}_k})\log\left(\frac{m^2_{\chi^0}}{\mu^2}\right) \right]~.
\eeqa

In order to obtain the final result we use Eq.~(\ref{main})
with Eqs.~(\ref{etanu}), (\ref{etaell}), and (\ref{etaW}). We get
\beqa
\label{fintoy}
\varepsilon =
\left(\eta^{W\dagger}+\eta^{W}-\eta^{\ell}-\eta^\nu\right)&=&
{g^{\prime}}^2q_{Y}^2 \sum_{k}U^{kj\,*}_{\tilde{\ell}} U^{ki}_{\tilde{\ell}}
\,\left(2{\mathcal I}^W_{k}-
{\mathcal I}^\ell_{k}-{\mathcal I}^\nu_{k}\right)\nonumber \\ &=&
{g^{\prime}}^2q_{Y}^2 \sum_{k}
U^{kj\,*}_{\tilde{\ell}} U^{ki}_{\tilde{\ell}}\,\left(2F^W_{k}-
F^\ell_{k}-F^\nu_{k}\right),
\eeqa
where we neglected $\mathcal{O}(\Lambda^{-2})$ effects.

We are now in a position to check the finiteness of the physical
amplitude.  Note that $\eta^\nu$, $\eta^\ell$, and $\eta^{W}$ contain
a $\log$-divergence.  $SU(2)_{L}$ gauge symmetry constraints the
coefficients of these two divergences to be equal to each other. We
can see that this is indeed the case by direct
inspection. $\varepsilon$ in Eq.~(\ref{fintoy}) depends only on the
functions $F^a$ that are independent of $\Lambda$. We can also check
that the result is independent of $\mu$, as it should be. For this note
that the sum, $2F^W_{k}- F^\ell_{k}-F^\nu_{k}$, is independent of
$\mu$.  While the above results are automatically achieved by each
off-diagonal term contributing to $\varepsilon$, as a result of the
GIM-mechanism, their validity for the diagonal components represents a
check of the correctness of the calculation.

Another important check is to make sure that in the limit of no EWSB,
there is no effect induced by the kinetic term because $SU(2)_{L}$
gauge symmetry makes it aligned with the $W$-interaction.
When sending $v_{EW}\rightarrow 0$, the charged sleptons become
degenerate with the sneutrinos, $m^2_{\tilde{l}_k}=m^2_{\tilde{\nu}_k}$.
In this limit, using (\ref{fintoy}) we learn that the relevant sum is
proportional to the identity
\beq
\varepsilon_{\alpha\beta} \propto
\left.\left(\eta^{W}_{\alpha\beta}+\eta^{W\dagger}_{\alpha\beta}-
\eta^{\nu}_{\alpha\beta}-\eta^{\ell}_{\alpha\beta}\right)\right|_{v_{EW}=0}
\propto U^{\dagger}_{\tilde{\ell}} U_{\tilde{\ell}} = \delta_{\alpha\beta}~.
\eeq
We learn that no flavor changing amplitude is induced thanks to the unitarity of $U_{\tilde{\ell}}$.

The fact that the effect vanishes for $v_{EW}=0$ can be used to get an approximate formula.
We can define the presumably small parameter
\beq
a_k \equiv 
\left(\frac{m^2_{\tilde{\nu}_k} - m^2_{\tilde{\ell}_k}}
{m^2_{\tilde{\ell}_k} + m^2_{\tilde{\nu}_k}}\right)\,,
\eeq
that vanishes for $v_{EW}=0$, and expand in $a_k$ to the leading order
\begin{align}
\label{allepsilon}
&\varepsilon_{\alpha\beta} =
\left(\eta^{W}+{\eta^{W}}^{\dagger}-\eta^{\nu}-
\eta^{\ell}\right)_{\alpha\neq\beta}
=\frac{g^{\prime\,2}q_{Y}^2}{16\pi^2}
U^{k\alpha\,*}_{\tilde{\ell}} U^{k\beta}_{\tilde{\ell}}
\sum_k\left[a_k^2 G_k +\mathcal{O}(a_k^3)\right]
\,,
\end{align}
where $G_k$ is a function of SUSY masses which does not vanish in the limit of $a_k \to 0$.
The dominant splitting between left handed sneutrinos and sleptons originates from the
$D$-terms and is flavor universal
\begin{equation}
\label{DtermsSplitting}
(m^2_{\tilde{\nu}_\alpha}-m^2_{\tilde{\ell}_\alpha})=
m_{Z}^2\cos^2\theta_{W}\cos(2\beta)~.
\end{equation}
As a result, in this toy model, $\varepsilon_{\mu\tau}$ can be estimated as
\begin{equation}
\varepsilon_{\mu\tau}
\sim
\frac{\alpha_{Y}}{4\pi}\cos^4\theta_{W}\cos^2(2\beta)
\left(\frac{m^2_{Z}}{\mbox{Max}[m^2_{\chi^0},m^2_{\tilde{\ell}}]}\right)^2
\delta^{L}_{\mu\tau}
\lesssim 10^{-6}\delta^{L}_{\mu\tau}\,.
\end{equation}
where we have defined the mass-insertion parameters
$\delta^{L}_{ij} = (M^2_L)_{ij}/\sqrt{(M^2_L)_{ii}(M^2_L)_{jj}}$, as usual.
Such values are well below the expected experimental resolutions 
even for $\delta^{L}_{\mu\tau} \sim 1$.

We discuss now the main differences arising in the cases where the $SU(2)$
breaking sources ii) and iii) are switched on.
In the case ii), the leading effect for $\varepsilon_{\mu\tau}$ reads
\begin{equation}
\varepsilon_{\mu\tau}
\sim
\frac{\alpha_{Y}}{4\pi}
\frac{m^2_{\tau}\left|A_{\tau}-\mu\tan\beta\right|^2}
{m^2_{\chi^0} m^2_{\tilde{\ell}}}\,
\delta^{L}_{\mu\tau}\,,
\end{equation}
where we picked up a double left-right mixing term for the third slepton generation.
In principle, $\varepsilon_{\mu\tau}$ could reach values even slightly above $10^{-4}$
for $m_{\tau}\mu\tan\beta/m^2_{\tilde{\ell}}\sim 1$; in practice the constraint from
$\tau\to\mu\gamma$ implies that $\varepsilon_{\mu\tau}< 10^{-5}$.

Finally, in the case iii) we get
\begin{equation}
\varepsilon_{\mu\tau}
\sim
\frac{\alpha_{2}}{4\pi}
\left|Z^{12}_{\pm}\right|^2
\delta^{L}_{\mu\tau}\,,
\end{equation}
where $Z^{12}_{\pm}$ are the mixing angles of the chargino mass matrix which read
\beq
Z^{12}_{+}\approx \frac{v_u M_2 + v_d\mu}{M^2_2-\mu^2}\,\qquad
Z^{12}_{-}\approx \frac{v_d M_2 + v_u\mu}{M^2_2-\mu^2}\,,
\eeq
where $\tan\beta=v_u/v_d$. We have explicitly checked that also in this
case $\varepsilon_{\mu\tau}< 10^{-5}$ after imposing the constraint from
$\tau\to\mu\gamma$.

We discuss now the box induced NSIs. These effects survive even in the
limit where all the $SU(2)$ breaking sources discussed above are negligible.
In particular, it turns out that the largest effects arise for light
sleptons/Winos and heavy Higgsino/Bino. The latter condition is necessary
to suppress ${\rm BR}(\tau\to\mu\gamma)$. As a result, it turns out that
\begin{equation}
\varepsilon_{\mu\tau}^{box}
\sim
\frac{\alpha_{2}}{4\pi}
\frac{m_{Z}^2}{\mbox{Max}[M^2_2, m^2_{\tilde{\ell}}]}\,
\delta^{L}_{\mu\tau}
\lesssim 10^{-3} \delta^{L}_{\mu\tau}
\,.
\end{equation}
As we will show in the numerical analysis, the box contribution
provides the dominant effect to $\varepsilon_{\mu\tau}$ that can reach
experimentally interesting levels while being still compatible with
the current bound on ${\rm BR}(\tau\to\mu\gamma)$. In fact, in the
most favorable situation where $M_2\sim m_{\tilde{\ell}}\ll\mu\sim
M_1$, one can find that
\begin{equation}
\left|\varepsilon_{\mu\tau}^{box}\right|
\approx
10^{-3} \sqrt{\frac{{\rm BR}(\tau\to\mu\gamma)}{10^{-7}}}\,,
\end{equation}
as we will confirm numerically.

\subsection{Scalar charged current}

\begin{figure}[ht]
\includegraphics[scale=1.]{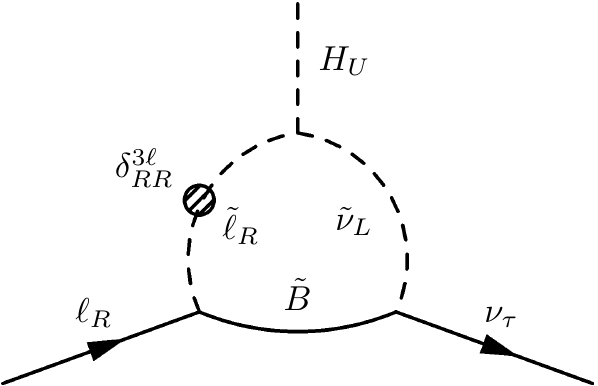}
\hskip 0.4 cm
\caption{A contribution to the effective  $\bar{\nu}_{\tau} \ell_R H^+$ coupling.}
\label{fig:higgs} 
\end{figure}

In the MSSM, Higgs mediated LFV effects are generated at the loop level, e.g.
see Fig.~\ref{fig:higgs}. In fact, given a source of non-holomorphic couplings,
and LFV among the sleptons, Higgs-mediated LFV is unavoidable~\cite{Babu:2002et}.

Starting from the model-independent parameterization for the flavor
violating couplings of a charged Higgs with leptons,
eq.~(\ref{Hln_LFV}), we specialize now to a SUSY scenario assuming
R-parity conservation to avoid tree level flavor changing effects.

Analysing the full expressions of such couplings (reported in the Appendix),  
we find that the field rotations $\delta L_m$ and $\delta R_m$ induce the 
dominant contributions for the effective Lagrangian of eq.~(\ref{Hln_LFV}).  
In fact, one can show that they are parametrically enhanced by a $\tan\beta$ 
factor $\delta L_m,\delta R_m \sim [\alpha_2/4\pi]\times \tan\beta$ compared 
to $\eta^{\ell,\nu}\sim \alpha_2/4\pi$ and $\eta^H \sim y_{\ell}\times[\alpha_2/4\pi]$.
 
Since the effects we are going to discuss can be relevant only if
$\tan\beta\gg 1$, it turns out that
\be\label{LH}
{\cal L}^{H^+}_{\rm eff}\simeq
\overline{\nu}_{jL}
\left[y^{\circ}_{\ell} -\delta L_m y^{\circ}_{\ell} + y^{\circ}_{\ell} \delta R_m \right]^{ji}
\ell_{iR} H^+ +{\rm H.c.}\,.
\ee
Retaining only the dominant $\tan\beta$ enhanced contributions in the corrections
to the lepton mass matrix, one has that $\delta m_\ell\simeq\eta_{m}^{\ell}$ and
therefore
\begin{eqnarray}
(\delta m_\ell)_{ij}
\simeq
   m^{\circ}_{\ell_i} {\eps} \tgb \delta_{ij}
   +
   \eps_{R} \tgb \delta_{R}^{ij} m^{\circ}_{\ell_j}
   +
   m^{\circ}_{\ell_i}\eps_{L} \tgb \delta_{L}^{ij}\,,
\end{eqnarray}
where $\eps,\eps_{L,R}$ are loop factors of order $\alpha_2/4\pi$
which depend on SUSY mass ratios and $t_\beta\equiv\tan\beta$.
Therefore, the rotation matrices can be determined explicitly from
eqs.~(\ref{eqn:dL}) and (\ref{eqn:dR}) and they read
\begin{eqnarray}
 \delta L_{m}^{3i}
 \simeq
\frac{\eps_{L}\tgb}{(1 + \eps\tgb)}\delta_{L}^{3i}\,,\qquad
 \delta R_{m}^{3i}
 \simeq
\frac{\eps_{R}\tgb}{(1 + \eps\tgb)}\delta_{R}^{3i} +
2\frac{m_{\ell_i}}{m_\tau}\frac{\eps_{L}\tgb}{(1 + \eps\tgb)}\delta_{L}^{3i}\,.
\end{eqnarray}
Finally, in the basis where $\nu_{L}\to L_m \nu_L$, the effective
Lagrangian for the $H^\pm$ couplings with leptons reads
\be\label{LH_eff}
{\cal L}^{H^+}_{\rm eff}
\simeq
\frac{g_2}{\sqrt{2}M_W}\frac{\tgb}{1 + \eps\tgb}~
\overline{\nu}_{jL}
\left[m_{\ell_i} \delta^{ji} + m_{\ell_i} \tgb \Delta^{ji}_{L} +
m_{\ell_j} \tgb \Delta^{ji}_{R} \right]
\ell_{iR} H^+ +{\rm H.c.}
\ee
where we have defined $\Delta^{ji}_{L(R)}\equiv\eps_{L(R)}\delta_{L(R)}^{ji}/(1+\eps\tgb)$.

An inspection of the above effective Lagrangian reveals that:
1) since the Yukawa operator is of dimension four, the quantities
$\Delta^{ji}_{L,R}$ depend only on ratios of soft SUSY masses, hence
avoiding SUSY decoupling.  Yet, the NP effects induced in physical
observables will decouple with the charged Higgs mass; 2) the loop
induced flavor violating couplings are enhanced by an extra $\tgb$
factor compared to the tree level flavor conserving couplings.
Therefore the loop suppression can be partially compensated if
$\Delta^{ji}_{L(R)}\tgb \sim 1$; and 3) the flavor violating
couplings $H^+\bar{\ell}\nu_{\tau}$ (with $\ell=e,\mu$) exhibit
a Yukawa enhancement factor $m_\tau/m_\ell$ compared to flavor
conserving couplings $H^+\bar{\ell}\nu_\ell$ when they are
induced by $\Delta^{ji}_{R}$.

Applying the above results to the model-independent parameterization
of eq.~(\ref{eq:eps_H}), we find
\beq
\varepsilon^{K}_{\mu\tau}
\approx
\left(\epsilon^{s\,*}_{\mu\tau}\right)^{K}=
-\left(\frac{m^{2}_{K}}{m^{2}_{H^\pm}}\right)
\left(
\Delta^{32}_{L} + \frac{m_{\tau}}{m_{\mu}}\Delta^{32}_{R}
\right)
\frac{\tgb^{3}}{(1+\eps_q\tgb)(1+\eps\tgb)}\,,
\label{eq:eps_H_susy}
\eeq
where $\eps_q$ is a non-holomorphic threshold correction stemming from the quark
sector typically of order $\eps_q\sim 10^{-2}$.

As seen in eq.~(\ref{eq:eps_H_pi_vs_k}), it turns out that
$\varepsilon^{\pi}_{\mu\tau}/\varepsilon^{K}_{\mu\tau}\approx 1/20$.
We notice that $\varepsilon^{\pi}_{\mu\tau}$ and $\varepsilon^{K}_{\mu\tau}$
show an enhanced sensitivity to sources of flavor violation in the right-handed 
slepton sector thanks to the Yukawa enhancement factor $m_{\tau}/m_{\mu}$.

In order to quantify the allowed size for $\varepsilon^{K,\pi}_{\mu\tau}$, we have
to impose the constraints arising from the charged lepton LFV decays. The most sensitive
probe of Higgs mediated effects is generally $\tau\to\ell_j\eta$~\cite{Sher:2002ew}
and the corresponding branching ratio reads
\beq
\frac{Br(\tau\rightarrow \mu\eta)}{Br(\tau\rightarrow \mu\bar{\nu_\nu}\nu_{\tau})}
\approx
10^{-2}
\left(\frac{|\Delta_{32}^{L}|^2 + |\Delta^{32}_{R}|^{2}}{m^{4}_{A}}\right)
\frac{\tgb^6}{|1+\eps_q\tgb|^2 |1+\eps\tgb|^2}\,,
\eeq
where $m_A$ is the pseudoscalar mass such that $m^{2}_{A}=m^{2}_{H^\pm}-M^2_W$
at tree level.
Imposing the experimental constraints from $Br(\tau\to\mu\eta)\lesssim 10^{-7}$,
it turns out that
\beq
\varepsilon_{\mu\tau}^{K}\lesssim 10^{-2}\,,
\qquad
\varepsilon_{\mu\tau}^{\pi}\lesssim 5\times 10^{-4}\,,
\eeq
where the above bounds arise for $|\Delta_{32}^{L}|\ll|\Delta^{32}_{R}|$.

Finally, let us mention that Higgs mediated LFV interactions also
induce lepton universality breaking effects in $P\to\ell\nu$
($\ell=e,\mu$)~\cite{Masiero:2005wr}. However, these effects can
only constrain $|\Delta^{31}_{R}|$ which is unrelated, in general,
with the relevant LFV term for NSIs, that is $|\Delta^{32}_{R}|$.

\subsection{Numerical analysis}

In this section, we provide the predictions for the NSI parameter
$\varepsilon_{\mu\tau}$ in the framework of the R-parity conserving 
MSSM with generic LFV soft breaking terms. The allowed values for
${\Im}(\varepsilon_{\mu\tau})$ are obtained after imposing the following
constraints: i) the data on flavor physics observables; ii) the mass
bounds from direct SUSY searches; iii) the requirement of a neutral
lightest SUSY particle; iv) the requirement of correct electroweak
symmetry breaking and vacuum stability; and v) the constraints from
electroweak precision observables.

Concerning NSI effects driven by the charged Higgs exchange, the most
stringent bounds come from the data on LFV and $B$-physics
observables. In particular, the processes $B\to X_s\gamma$,
$B\to\tau\nu$ and $B\to D\tau\nu$ are known to be the most powerful
probes of new charged scalar currents. In principle, also the process
$B_{s,d}\to\mu^+\mu^-$ shows an enhanced sensitivity to extended Higgs
sectors. However, since the loop-induced flavor changing coupling
$H\bar{b}s(d)$ (with $H = H^0,A^0$) depends on the details of the soft
sector, to be conservative, we do not impose here the
(model-dependent) constraint from ${\rm BR}(B_{s,d}\to\mu^+\mu^-)$.

The bounds from ${\rm BR}(B\to X_s\gamma)$ have been obtained
employing the SM prediction at the NNLO of Ref.~\cite{bsgth}, ${\rm
BR}(B\to X_s\gamma;E_\gamma>1.6~{\rm GeV})^{\rm SM}=(3.15 \pm 0.23)
\times 10^{-4}$, combined with the experimental
average~\cite{hfag,belle,babar} ${\rm BR}(B\to X_s
\gamma;E_\gamma>1.6~{\rm GeV}))^{\rm exp}=(3.55\pm 0.24)\times
10^{-4}$.  As for the SUSY contributions, we use the calculation
of Ref.~\cite{bsgamma} assuming decoupled gluinos and squarks.
For $B\to\tau\nu$, we use the current world average ${\rm
BR}(B\to\tau\nu)_{\rm exp}=(1.73 \pm 0.35)\times
10^{-4}$~\cite{btaunuexp}, the SM prediction $(1.10 \pm 0.29)\times
10^{-4}$~\cite{abgps} (see also~\cite{Bona:2009cj}) and the NP
contributions of Ref.~\cite{hou}.

Finally, the NP sensitivity of $B\to D\tau\nu$ can be better exploited
normalizing it to ${\rm BR}(B \to D \tau \nu)/{\rm BR}(B \to D \ell
\nu)$ where $\ell=e,\mu$~\cite{bdtaunu_old,itoh}. We use the world 
average $(49\pm 10)\%$~\cite{bdtaunuexp} and the theoretical prediction 
of Ref.~\cite{mescia}.  

In our numerical analysis we impose all the above constraints at the $2\sigma$ C.L..

\begin{figure}[tb]
\begin{center}
\includegraphics[scale=0.37]{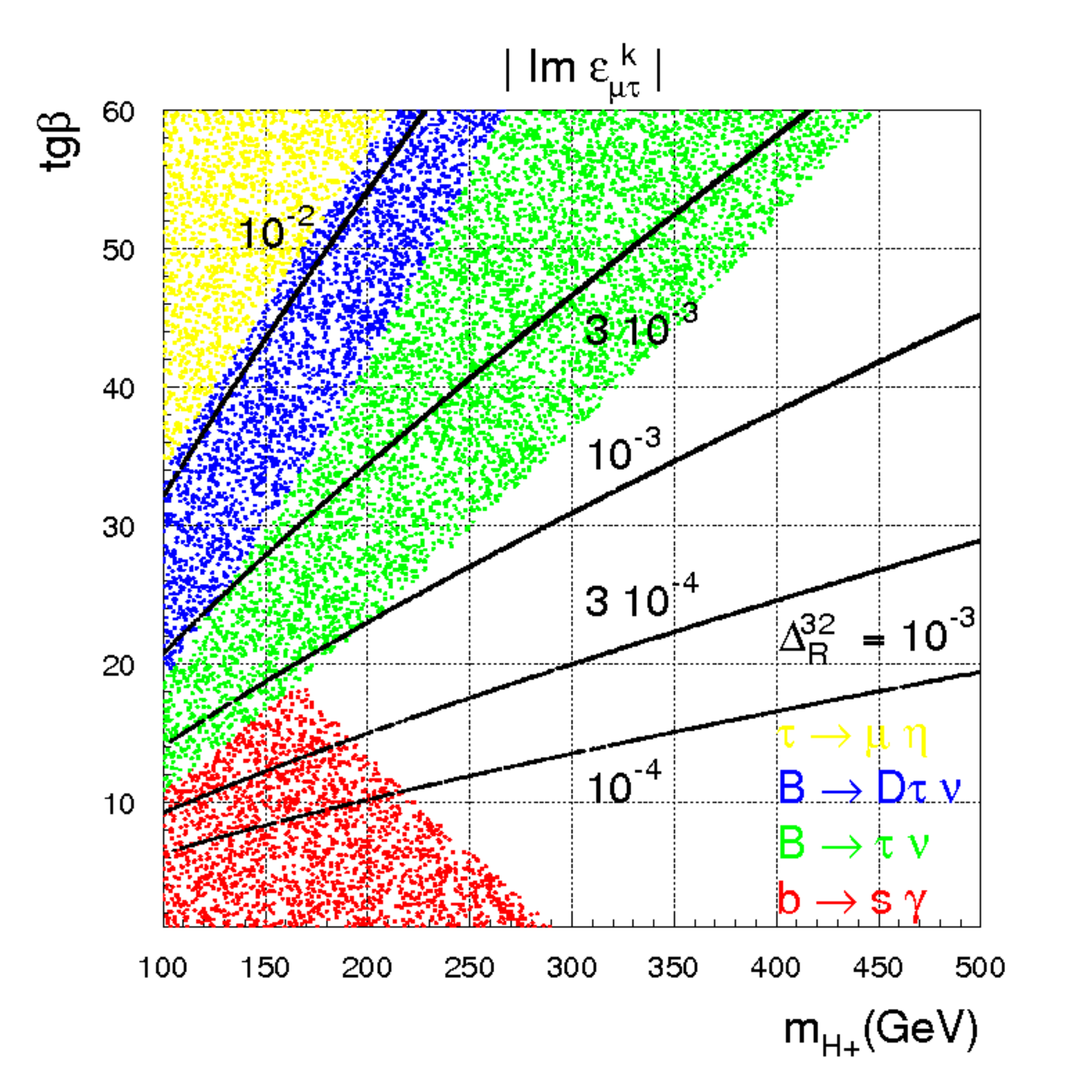}~
\includegraphics[scale=0.37]{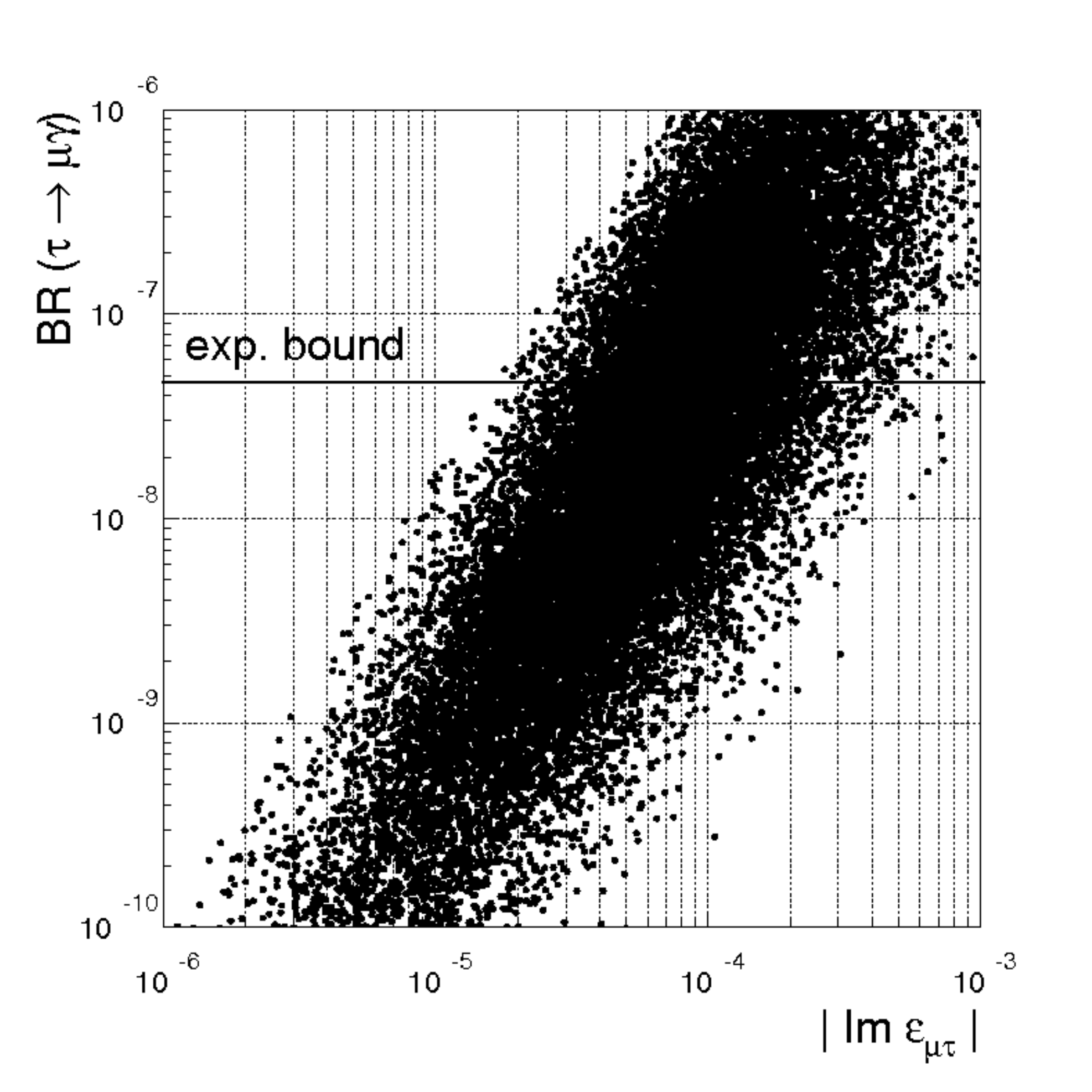}
\caption{Left: NSIs in the process $K\to\ell\nu$ induced by Higgs mediated effects.
Right: NSIs in the process $\mu\to e\nu\bar\nu$ induced by $W$-penguin and gaugino/slepton
boxes. See the text for details.}
\label{fig:nsi_H}
\end{center}
\end{figure}

In Fig.~(\ref{fig:nsi_H}) on the left, we show the values attained by
$|{\rm Im}\varepsilon_{\mu\tau}^K|$, see eq.~(\ref{eq:eps_H_susy}), in the
$\tan\beta-M_{H^\pm}$ plane setting the LFV parameter $|\Delta^{32}_{R}|=10^{-3}$
(varying $\Delta^{32}_{R}$, $|{\rm Im}\varepsilon_{\mu\tau}^K|$ would rescale
according to $|\Delta^{32}_{R}|/10^{-3}$). The red, green, glue and yellow
regions are excluded by the current bounds on $B\to X_s\gamma$, $B\to\tau\nu$,
and $B\to D\tau\nu$ and $\tau\to\mu\eta$, respectively.

As shown by fig.~(\ref{fig:nsi_H}), $|{\rm Im}\varepsilon_{\mu\tau}^K|$ can vary in
the range $(10^{-4},10^{-2})$ for $\tan\beta\leq 60$ and $M_{H^\pm}\leq 500$~GeV.
The corresponding values for $|{\rm Im}\varepsilon_{\mu\tau}^{\pi}|$ can be obtained
by $|{\rm Im}\varepsilon_{\mu\tau}^{\pi}|/|{\rm Im}\varepsilon_{\mu\tau}^{K}|\approx 1/20$.

We now discuss the NSIs as induced by the $V-A$ charged current via the one loop exchange
of gauginos/sleptons. As discussed in the above section, the dominant effect to
${\rm Im}\varepsilon_{\mu\tau}$ arises from the box contributions. In fig.~(\ref{fig:nsi_H}) 
on the right, we show the correlation between ${\rm BR}(\tau\to\mu\gamma)$ vs.
$|{\rm Im}\varepsilon_{\mu\tau}|$ in the case where the neutrino source is provided by 
the muon decay $\mu\to e\nu_\tau\bar{\nu}_e$.
We have assumed heavy squarks implying negligible NSIs at the detector level. In this limit
also NSIs for the production process $P\to\mu\nu_{\tau}$ are suppressed.
Moreover, as the largest effects for ${\rm Im}\varepsilon_{\mu\tau}$ are obtained for light
sleptons/Winos and heavy Higgsino/Bino (to keep under control ${\rm BR}(\tau\to\mu\gamma)$),
we employ the following scan over the SUSY input parameters: $M_2,m_{\tilde{\ell}}\leq 1$~TeV,
$\mu, M_1> 500$~GeV and $3<\tan\beta<10$.

As shown by fig.~(\ref{fig:nsi_H}), $|{\rm Im}\varepsilon_{\mu\tau}|$ can reach experimentally 
interesting values $|{\rm Im}\varepsilon_{\mu\tau}|\lesssim 3-4\times 10^{-4}$ and this would 
unambiguously imply a lower bound for ${\rm BR}(\tau\to\mu\gamma)$ quite close
to the current bound.

\section{Discussion and conclusion}
\label{conclusions}

The idea that neutrino oscillations can probe NSIs is very attractive.
In theory such experiments are sensitive to any form of new physics
that makes the produced and detected neutrinos non-orthogonal.
Such non-orthogonality, parameterized by $\varepsilon_{\alpha\beta}$,
may come from new tree level interactions, new heavy neutrinos, or one 
loop effects that modify the couplings of the $W$ boson to the leptons.

In this work we presented a general framework that allows one to
extract in a consistent way the physical $\varepsilon$ arising at the
loop level either from the $V-A$ or scalar charged currents. We show
how $\varepsilon$ can be obtained from the various loop amplitudes
which include vertex corrections, wave function renormalizations,
mass corrections as well as box diagrams. 

As an illustrative example, we discussed NSIs in the R-parity conserving
MSSM with new LFV sources in the soft sector.

We argued that, in general, the size of one-loop NSIs is quite small,
$\varepsilon\approx\mathcal{O}(10^{-3})$. To be observed, such small
numbers require very precise measurements of the neutrino appearance
probability as a function of $L/E$. We hope that such measurements
will be possible in the next generation of neutrino oscillation
experiments.

\acknowledgments
The work of BB, YG and IN is supported by the NSF grant PHY-0757868.
The work of PP has been partially supported by the German
`Bundesministerium f\"ur Bildung und Forschung' under contract 05H09WOE.

%
\section{Appendix}
In the following, we provide the full analytical expressions for the self-energies
and vertex corrections relevant for NSIs in the R-parity conserving MSSM.
For the Feynman rules, we closely follow the notation of Ref.~\cite{Rosiek:1989rs}.
%

The lepton self-energies read
\bea
-(4\pi)^2 \left(\eta^{\nu}_{VL}\right)^{IJ}&=&
L^{Iki}_{\nu LC} L^{Jki\,*}_{\nu LC} B_{1}(m_{L_k},m_{C_i}) +
L^{Iki}_{\nu\tilde{\nu} N} L^{Jki\,*}_{\nu\tilde{\nu} N}
B_{1}(m_{\tilde{\nu}_k},m_{N_i})
\\
-(4\pi)^2 \left(\eta^{\ell}_{VL}\right)^{IJ}&=&
L^{Iki}_{eLN} L^{Jki\,*}_{eLN}B_{1}(m_{L_k},m_{N_i})+
L^{Iki}_{e\tilde{\nu} C} L^{Jki\,*}_{e\tilde{\nu} C}
B_{1}(m_{\tilde{\nu}_k},m_{C_i})
\\
-(4\pi)^2 \left(\eta^{\ell}_{VR}\right)^{IJ}&=&
R^{Iki}_{eLN} R^{Jki\,*}_{eLN}B_{1}(m_{L_k},m_{N_i})+
R^{Iki}_{e\tilde{\nu} C} R^{Jki\,*}_{e\tilde{\nu} C}
B_{1}(m_{\tilde{\nu}_k},m_{C_i})
\\
(4\pi)^2 \left(\eta^{\ell}_{mL}\right)^{IJ}&=&
-L^{Iki}_{eLN} R^{Jki\,*}_{eLN}B_{0}(m_{L_k},m_{N_i})
-L^{Iki}_{e\tilde{\nu} C} R^{Jki\,*}_{e\tilde{\nu} C}
B_{0}(m_{\tilde{\nu}_k},m_{C_i})\,.
\eea
The vertex corrections relevant for $W\ell\nu$ are
\bea
(4\pi)^2 \left(\eta^{W}\right)^{IJ}
\!\!\!&=&
\frac{1}{2}
L^{Jkj\,*}_{\nu\tilde{\nu}N} L^{Iij}_{eLN} Z^{Lk\,*}_{\nu} Z^{Li\,*}_{L}
\left[B_{0}(m_{L_i},m_{\tilde{\nu}_k}) + \frac{1}{2} + m^{2}_{N_j}
C_{0}(m_{N_j},m_{L_i},m_{\tilde{\nu}_k}) \right] +
\nonumber\\
&+&L^{Jki\,*}_{\nu LC} L^{Ikj}_{eLN}
\bigg[ \sqrt{2} L^{ji}_{wCN} m_{C_i} m_{N_j} C_{0}(m_{L_k},m_{C_i},m_{N_j})+
\nonumber\\
&-&\frac{1}{\sqrt{2}} R^{ji}_{wCN}
\left(B_{0}(m_{C_i},m_{N_j}) - \frac{1}{2} + m^{2}_{L_k} C_{0}(m_{L_k},m_{C_i},m_{N_j}) \right)
\bigg]
\nonumber\\
&+&
L^{Jkj\,*}_{\nu\tilde{\nu}N} L^{Iki}_{e\tilde{\nu}C}
\bigg[ -\sqrt{2} R^{ji}_{wCN} m_{C_i} m_{N_j} C_{0}(m_{\tilde{\nu}_k},m_{C_i},m_{N_j})+
\nonumber\\
&+&
\frac{1}{\sqrt{2}} L^{ji}_{wCN}
\left(B_{0}(m_{C_i},m_{N_j}) - \frac{1}{2} + m^{2}_{\tilde{\nu}_k}
C_{0}(m_{\tilde{\nu}_k},m_{C_i},m_{N_j}) \right)
\bigg]\,.
\eea
The vertex corrections relevant for $H\ell\nu$ are
\bea
\left(4\pi\right)^2 \left(\eta^{H}\right)^{IJ} &=&
-V_{\tilde{\nu}LH}^{ml} L_{\nu\tilde{\nu}N}^{Jmn\star}
R_{eLN}^{Iln} m_{N_n} C_0(m_{N_n},m_{\tilde{\nu}_m},m_{L_l})
\nonumber\\
&+& L_{\nu LC}^{Jnm\star} R_{eLN}^{Inl}
\left[L_{NCH}^{lm} C_2(m_{L_n}^2,m_{C_m}^2,m_{N_l}^2)\right.
\nonumber\\
&&\phantom{aaaaaaaaa}
-\left. R_{NCH}^{lm}
m_{C_m} m_{N_l} C_0(m_{L_n}^2,m_{C_m}^2,m_{N_l}^2)\right]
\nonumber\\
&+& L_{\nu\tilde{\nu}N}^{Jnl\star} R_{e\tilde{\nu}C}^{Inm}
\left[L_{NCH}^{lm} C_2(m_{\tilde{\nu}_n}^2,m_{N_l}^2,m_{C_m}^2) \right.\nonumber\\
&&\phantom{aaaaaaaaa}
-\left. R_{NCH}^{lm} m_{N_l} m_{C_m} C_0(m_{\tilde{\nu}_n}^2,m_{N_l}^2,m_{C_m}^2)
\right]\,.
\eea
%


The gaugino/slepton box diagrams relevant for the process
$\mu\to e\nu_\tau\bar{\nu_e}$ read
\bea
-(4\pi)^2 \epsilon^{\rm box}_{IJ}
&=&
L^{Ikj}_{\ell LN} L^{Jki\,*}_{\nu LC}
L^{eli}_{\nu LC} L^{elj\,*}_{\ell LN}
D_{2}(m_{L_k},m_{L_l},m_{C_i},m_{N_j})
\nonumber\\
&+&
L^{IKi}_{\ell \tilde{\nu} C} L^{JKj\,*}_{\nu \tilde{\nu} N}
L^{eLj}_{\nu\tilde{\nu}N} L^{eLi\,*}_{\ell \tilde{\nu} C}
D_{2}(m_{\tilde{\nu}_K},m_{\tilde{\nu}_L},m_{C_i},m_{N_j})
\nonumber\\
&+&
{1\over2}
L^{Ikj}_{\ell LN} L^{eKj}_{\nu \tilde{\nu} N}
L^{Jki\,*}_{\nu LC} L^{eKi\,*}_{\ell \tilde{\nu} C} m_{C_i} m_{N_j} 
D_{0}(m_{L_k},m_{\tilde{\nu}_K},m_{C_i},m_{N_j})
\nonumber\\
&+&
{1\over2}
L^{IKi}_{\ell \tilde{\nu} C} L^{eki}_{\nu LC}
L^{JKj\,*}_{\nu \tilde{\nu} N} L^{ekj\,*}_{\ell LN} m_{C_i} m_{N_j} 
D_{0}(m_{L_k},m_{\tilde{\nu}_K},m_{C_i},m_{N_j})\,.
\eea
%
%
The box diagrams generated by the gaugino/slepton(squark) exchange contributing
to the production process $P\to\mu\nu_\alpha$ ($P=\pi,K$) and to the detection
process read
\bea
-(4\pi)^2 \epsilon^{\rm box}_{IJ}
&=&
L^{Jkj\,*}_{\ell LN} L^{Iki}_{\nu LC}
L^{dli\,*}_{uDC} L^{dlj}_{dDN}
D_{2}(m_{\tilde{\ell}_k},m_{\tilde{d}_l},m_{C_i},m_{N_j})
\nonumber\\
&+&
L^{JKi\,*}_{\ell\tilde{\nu} C} L^{IKj}_{\nu\tilde{\nu} N}
L^{dLj\,*}_{uUN} L^{dLi}_{dUC}
D_{2}(m_{\tilde{\nu}_K},m_{\tilde{u}_L},m_{C_i},m_{N_j})
\nonumber\\
&+&
{1\over2}
L^{Jkj\,*}_{\ell LN} L^{dKj\,*}_{uUN}
L^{Iki}_{\nu LC} L^{dKi}_{dUC} m_{C_i} m_{N_j} 
D_{0}(m_{L_k},m_{\tilde{u}_K},m_{C_i},m_{N_j})
\nonumber\\
&+&
{1\over2}
L^{JKi\,*}_{\ell\tilde{\nu} C} L^{dki\,*}_{uDC}
L^{IKj}_{\nu\tilde{\nu}N} L^{dkj}_{dDN} m_{C_i} m_{N_j} 
D_{0}(m_{\tilde{d}_k},m_{\tilde{\nu}_K},m_{C_i},m_{N_j})\,.
\eea
The expressions for the loop functions appearing in the above amplitudes read
\bea
B_{0}(m_1,m_2)= \frac{1}{\varepsilon} + 1 - \frac{1}{m^2_1 - m^2_2}
\left[m^2_1~{\rm log}\frac{m^2_1}{\mu^2} - m^2_2~{\rm log}\frac{m^2_2}{\mu^2} \right]\,,
\eea
\bea
B_{1}(m_1,m_2)=
-\frac{1}{2}
\left[
\frac{1}{\varepsilon} + 1 - {\rm log}\frac{m^2_2}{\mu^2} +
\left(\frac{m^2_1}{m^2_1 - m^2_2}\right)^2 {\rm log}\frac{m^2_2}{m_1^2}+
\frac{1}{2}\frac{m^2_1 + m^2_2}{m^2_1 - m^2_2}
\right]\,,
\eea
\bea
C_{0}(m_1,m_2,m_3)=
\frac{1}{m^2_2 - m^2_3}
\left[
\frac{m^2_2}{m^2_1-m^2_2}~{\rm log}\frac{m^2_2}{m^2_1}-
\frac{m^2_3}{m^2_1-m^2_3}~{\rm log}\frac{m^2_3}{m^2_1}
\right]\,,
\eea
\bea
C_2(m_1,m_2,m_3)= \frac{1}{\varepsilon} + 1 + \log{m_1^2\over\mu^2}
+{m_2^4\log{m_2^2 / m_1^2}\over(m_1^2-m_2^2)(m_3^2-m_2^2)}
+{m_3^4 \log{m_3^2 / m_1^2} \over(m_1^2-m_3^2)(m_2^2-m_3^2)}\,,
\eea
\bea
 D_0(m_1,m_2,m_3,m_4) &=&
\frac{m_1^2 \log{m_1^2}}{(m_4^2\!-\!m_1^2)(m_3^2\!-\!m_1^2)(m_2^2\!-\!m_1^2)}
\nonumber\\
&+&
\{1 \leftrightarrow 2\} \!+\! \{1 \leftrightarrow 3\} \!+\! \{1 \leftrightarrow 4\}\,,
\eea
\bea
D_2(m_1,m_2,m_3,m_4) &=&
\frac{1}{4}\frac{m_1^4 \log{m_1^2}}{(m_4^2\!-\!m_1^2)(m_3^2\!-\!m_1^2)(m_2^2\!-\!m_1^2)}
\nonumber\\
&+&\{1 \leftrightarrow 2\} \!+\! \{1 \leftrightarrow 3\} \!+\! \{1 \leftrightarrow 4\}\,.
\eea


\end{document}